\documentclass[a4paper,fleqn,usenatbib]{mnras}

\usepackage{newtxtext,newtxmath}

\usepackage[T1]{fontenc}
\usepackage{ae,aecompl}

\usepackage{graphicx}	
\usepackage{amsmath}	
\usepackage{amssymb}	

\usepackage{longtable,lscape}
\usepackage{rotating}
\usepackage{multirow,multicol}

\title[SN 2018zd]{SN 2018zd: An Unusual Stellar Explosion as Part of the Diverse Type II Supernova Landscape}
\author[J.-J. Zhang et al.]{Jujia Zhang,$^{1,2,3,4}$\thanks{E-mail: jujia@ynao.ac.cn}
Xiaofeng Wang,$^{5,6}$
J\'{o}zsef Vink\'{o}$^{7,8,9}$
Qian Zhai,$^{1,2,3,4}$
Tianmeng Zhang,$^{10,11}$ \newauthor
Alexei V. Filippenko,$^{12,13}$
Thomas G. Brink,$^{12}$
WeiKang Zheng,$^{12}$
{\L}ukasz Wyrzykowski,$^{14}$\newauthor
Przemys{\l}aw Miko{\l}ajczyk,$^{14}$
Fang Huang,$^{15}$
Liming Rui,$^{5}$
Jun Mo,$^{5}$
Hanna Sai,$^{5}$\newauthor
Xinhan Zhang,$^{5}$
Huijuan Wang,$^{10,11}$
James M. DerKacy,$^{16}$
Eddie Baron,$^{16}$
K. S\'arneczky,$^{7}$\newauthor
A. B\'odi,$^{7,18}$
G. Cs\"ornyei,$^{7,8}$ 
O. Hanyecz,$^{7}$ 
B. Ign\'acz,$^{7}$ 
Cs. Kalup,$^{7,8,18}$ 
L. Kriskovics,$^{7,8}$\newauthor
R. K\"onyves-T\'oth,$^{7,8}$ 
A. Ordasi,$^{7}$ 
A. P\'al,$^{7,8,17}$ 
\'A. S\'odor,$^{7,18}$
R. Szak\'ats,$^{7}$
K. Vida,$^{7,8,18}$\newauthor
G. Zsidi$^{7,8,19}$
\\
$^{1}$Yunnan Observatories (YNAO), Chinese Academy of Sciences (CAS), Kunming, 650216, China\\
$^{2}$Chinese Academy of Sciences South America Center for Astronomy, National Astronomical Observatories, CAS, Beijing 100012, China\\
$^{3}$Key Laboratory for the Structure and Evolution of Celestial Objects,CAS, Kunming, 650216, China\\
$^{4}$Center for Astronomical Mega-Science, CAS, 20A Datun Road, Chaoyang District, Beijing, 100012, China\\
$^{5}$Physics Department and Tsinghua Center for Astrophysics (THCA), Tsinghua University, Beijing, 100084, China\\
$^{6}$Beijing Planetarium, Beijing Academy of Science and Technology, Beijing, 100044, China\\
$^{7}$Konkoly Observatory, CSFK, Konkoly-Thege M. ut 15-17, Budapest, 1121, Hungary\\
$^{8}$ELTE E\"otv\"os Lor\'and University, Institute of Physics, P\'azm\'any P\'eter s\'et\'any 1/A, Budapest, 1117, Hungary\\
$^{9}$Department of Optics and Quantum Electronics, University of Szeged, D\'om t\'er 9, Szeged, 6720, Hungary\\
$^{10}$Key Laboratory of Optical Astronomy, National Astronomical Observatories, Chinese Academy of Sciences, Beijing, 100012, China\\
$^{11}$School of Astronomy and Space Science, University of Chinese Academy of Sciences, 101408, Beijing, China\\
$^{12}$Department of Astronomy, University of California, Berkeley, CA 94720-3411, USA\\
$^{13}$Miller Senior Fellow, Miller Institute for Basic Research in Science, University of California, Berkeley, CA  94720, USA\\
$^{14}$Astronomical Institute University of Wroc{\l}aw, M. Kopernika 11, 51-622 Wroc{\l}aw, Poland\\
$^{15}$Department of Astronomy, Shanghai Jiao Tong University, Shanghai, 200240, China\\
$^{16}$Homer L. Dodge Department of Physics and Astronomy, University of Oklahoma, Norman, OK 73019, USA\\
$^{17}$ELTE E\"otv\"os Lor\'and University, Department of Astronomy, P\'azm\'any P\'eter s\'et\'any 1/A, Budapest, 1117, Hungary\\
$^{18}$MTA CSFK Lend\"ulet Near-Field Cosmology Research Group, Konkoly Observatory, Budapest, 1121, Hungary \\
$^{19}$European Southern Observatory, Karl-Schwarzschild-Strasse 2, 85748 Garching bei M{\"u}nchen, Germany
}

\date{Accepted 2020 July 28. Received 2020 April 11; in original form 2020 April 11}

\pubyear{xxxx}

\begin{document}
\label{firstpage}
\pagerange{\pageref{firstpage}--\pageref{lastpage}}
\maketitle

\begin{abstract}
We present extensive observations of SN 2018zd covering the first $\sim450$\,d after the explosion. This SN shows a possible shock-breakout signal $\sim3.6$\,hr after the explosion in the unfiltered light curve, and prominent flash-ionisation spectral features within the first week. The unusual photospheric temperature rise (rapidly from $\sim 12,000$\,K to above 18,000\,K) within the earliest few days suggests that the ejecta were continuously heated. Both the significant temperature rise and the flash spectral features can be explained with the interaction of the SN ejecta with the massive stellar wind ($0.18^{+0.05}_{-0.10}\, \rm M_{\odot}$), which accounts for the luminous peak ($L_{\rm max} = [1.36\pm 0.63] \times 10^{43}\, \rm erg\,s^{-1}$) of SN 2018zd.  The luminous peak and low expansion velocity ($v \approx 3300$ km s$^{-1}$) make SN 2018zd to be like a member of the LLEV (luminous SNe II with low expansion velocities) events originated due to circumstellar interaction. 
The relatively fast post-peak decline allows a classification of SN 2018zd as a transition event morphologically linking SNe~IIP and SNe~IIL. In the radioactive-decay phase, SN 2018zd experienced a significant flux drop and behaved more like a low-luminosity SN~IIP both spectroscopically and photometrically.  This contrast indicates that circumstellar interaction plays a vital role in modifying the observed light curves of SNe~II. Comparing nebular-phase spectra with model predictions suggests that SN 2018zd arose from a star of  $\sim 12\,\rm M_{\odot}$.  
Given the relatively small amount of $^{56}$Ni ($0.013 - 0.035 \rm M_{\odot}$),  the massive stellar wind,  and the faint X-ray radiation, the progenitor of SN 2018zd could be a massive asymptotic giant branch star which collapsed owing to electron capture.   
 \end{abstract}



\section{Introduction}

Type II supernovae (SNe~II) are  hydrogen-rich core-collapse events that are observationally divided into Type IIP (SNe IIP), Type IIL (SNe IIL), Type IIn (SNe IIn), and Type IIb (SNe IIb) \citep[for reviews, see, e.g.,][]{Filippenko97,Gal-Yam17}. Among them, SNe IIP belong to the most abundant subclass, which is characterised by a relatively constant  optical  luminosity plateau lasting for about three months ($\sim 100$\,d) after the explosion, followed by a rapid drop to the radioactive tail. The thermalisation of the initial shock wave and the recombination of the ionised hydrogen provide sources to power  the plateau light curve \citep{Popov93}. SNe IIL were named after their linear (in magnitudes) light-curve decay that starts soon after peak brightness. Based on the understanding of SNe~IIP, the absence of a plateau in the light curves of SNe~IIL might suggest less energy input at a similar phase, while the absence of an abrupt drop in brightness after a few months might indicate a lower-mass H envelope in their progenitors. However, the higher peak luminosity of SNe~IIL ($\sim 1.5$\,mag brighter than that of SNe~IIP, on average; \citealp{Patat93,Patat94,Li11,Anderson14,Faran14b,Sanders15}) disfavours this hypothesis. The continuous range of early-time light-curve slopes found among SNe~II \citep[e.g.,][]{Anderson14,Sanders15}, and the observed transitional events (e.g., SN 2013ej, \citealp{Huang15}; SN 2013by, \citealp{Valenti15}) tends to suggest a continuous progenitor population for SNe~IIP and SNe~IIL. 

A fraction of SNe~II exhibit signatures of ejecta interaction with circumstellar material (CSM) produced by mass loss in massive stars, especially at their late phases before core collapse. The events that present prominent interaction signatures are classified as SNe~IIn \citep{Schlegel90}.  The observed diversity of SNe~IIn indicates that the circumstellar environments around their progenitors are complicated.  The duration of interaction has large scatter, spanning from a few days (e.g., SN 2013fs, \citealp{Yaron17}), a few weeks (e.g., SN 1998S, \citealp{Fassia00,Leonard00}),  a few months (e.g., SN 2010jl, \citealp{Zhang12,Fransson14}),  to even a few years (e.g., SN 2015da, \citealp{Tartaglia20}).

The spectra of SNe~IIb are similar to those of the SNe~IIP and IIL (with strong  lines of H) near maximum light, but in the following weeks they usually metamorphoze into SNe~Ib \citep{Filippenko88,Filippenko+93}. We do not discuss SNe~IIb in this paper because they show more similarities to hydrogen-poor events (SNe~Ibc, \citealp{Arcavi12,Stritzinger18}).  

The physical origin of SNe~IIP and IIL, and their connections with SNe~IIn regarding the physical characteristics of their progenitor scenarios and explosion properties, have been long-standing issues. There are pieces of evidence indicating that the diversity between SNe~IIP and  IIL can be partly explained with short-lived interaction \citep{Valenti15,Morozova17}. We present optical and ultraviolet (UV) data for the core-collapse SN 2018zd, obtained through an observational campaign that lasted for about 450\,d with several telescopes. This SN shows a series of interaction signatures in both spectra and light curves, as well as a large flux drop before entering the radioactive-decay phase.

Observations and data reduction are described in Section \ref{sect:obs}. The UV and optical light curves are presented in Section \ref{sect:LC}, while the spectral evolution is shown in Section \ref{sect:SP}. In Section \ref{sect:Dis}, we discuss the bolometric light curve, explosion parameters, progenitor properties, velocity evolution,  the position of this SN in the SNe~II family, and the possibility of originating from the electron-capture channel. A brief conclusion is given in Section \ref{sect:sum}.

\begin{figure}
\centering
\includegraphics[width=8cm,angle=0]{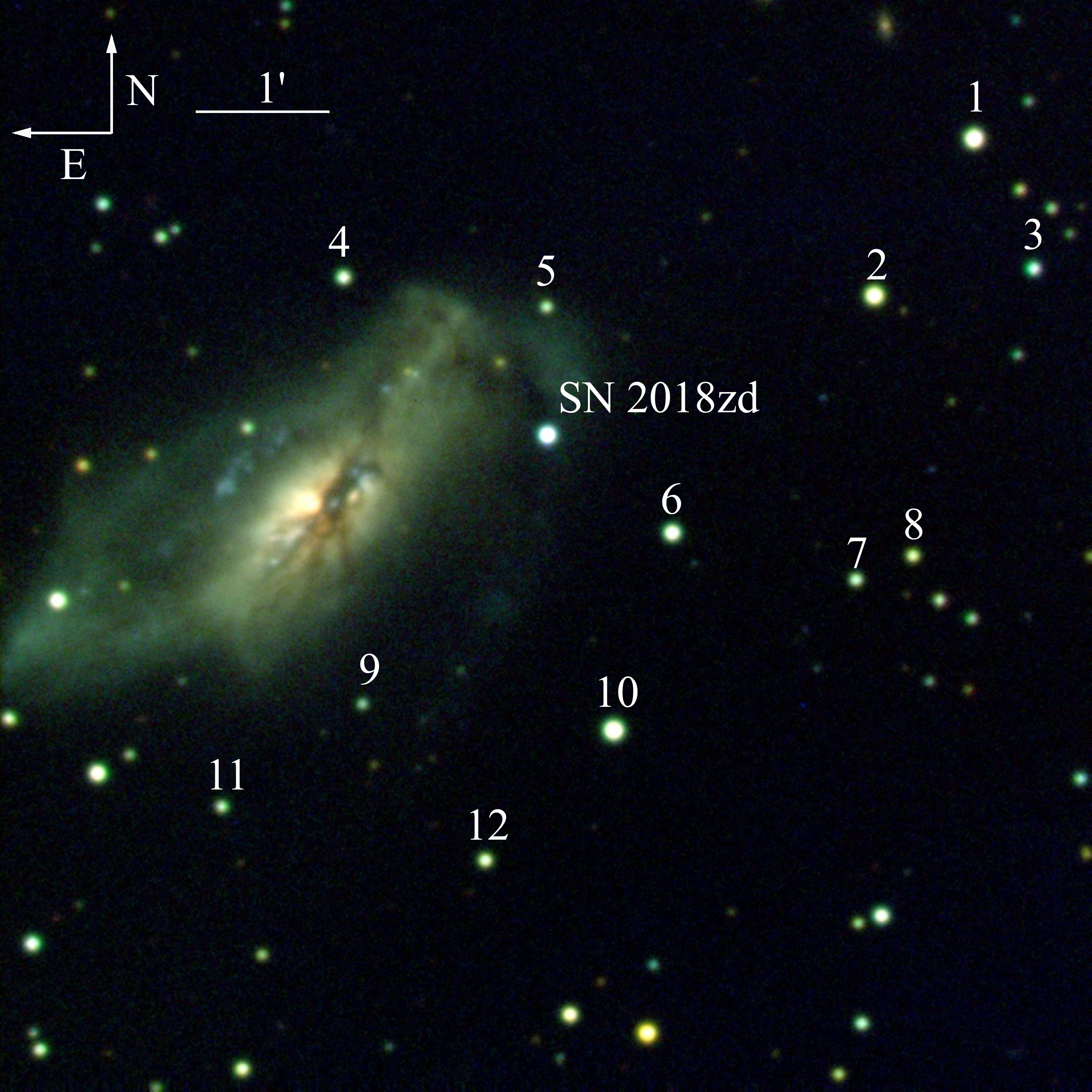}
 \caption{Finder chart of SN 2018zd in NGC 2146. The local reference stars listed in Table \ref{Tab:Photo_stand} are marked with a nearby number. This chart was combined by the $UBV$-band images taken with the 2.4\,m LJT and YFOSC on March 7, 2018, with a pixel size of 0.$\arcsec$57 in the Bin2 readout mode. The average FWHM is 2$\arcsec$.20 because of the high airmass ($\sim 1.90$) during the observation.}
\label{<img>}
\end{figure}

\begin{figure*}
\centering
\includegraphics[width=15cm,angle=0]{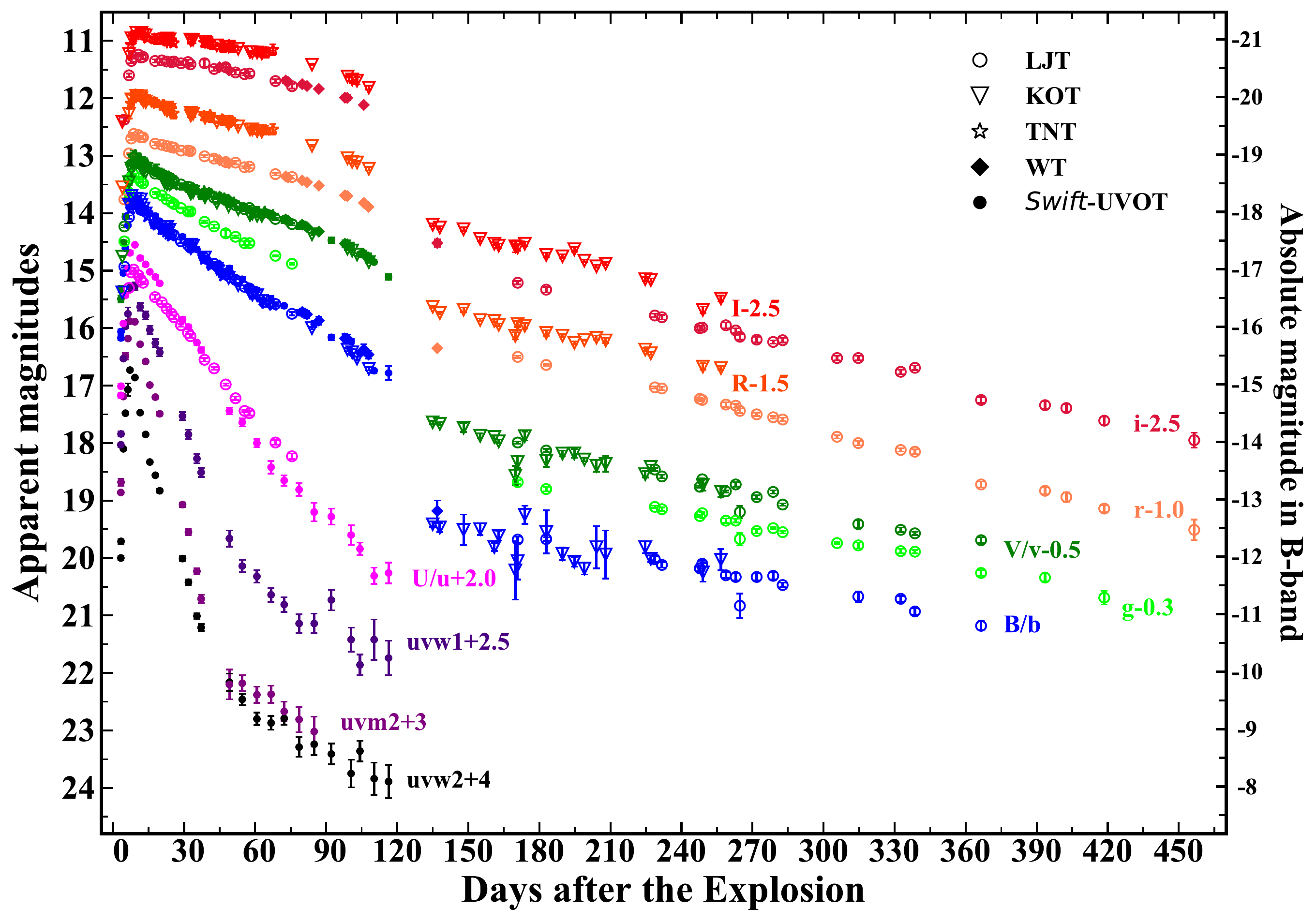}
 \caption{UV-optical light curves of SN 2018zd obtained with the 2.4\,m LJT, the 0.8\,m KOT, the 0.8\,m TNT, the 0.6\,m WT, and the {\it Swift} UVOT. All measurements are shifted vertically for better visibility.}
\label{<LC>}
\end{figure*}

\section{Observations} 
\label{sect:obs}

Koichi Itagaki reported the discovery of the transient AT 2018zd in the nearby SB(s)ab galaxy NGC 2146 on March 02.49, 2018 (UT dates are used throughout this paper). This transient was discovered at 17.8\,mag in an unfiltered image obtained with a 0.5\,m reflector at the Takamizawa station of Japan\footnote{http://www.k-itagaki.jp}. Its J2000 coordinates are $\alpha = 06^{\rm h}18^{\rm m}03.18^{\rm s}$ and $\delta = 78^\circ 22\arcmin 00\farcs90$, which is $104\farcs5$ west and $35\farcs6$ north of the centre of the host galaxy.  

The earliest spectroscopic observations of AT 2018zd were initiated at Li-Jiang Observatory of Yunnan Observatories (YNAO) with the Li-Jiang 2.4\,m telescope (hereafter LJT; \citealp{Fan15}) equipped with the YFOSC (Yunnan Faint Object Spectrograph and Camera; \citealp{Wang19}) only 3\,hr after the discovery. However, the noisy spectrum precluded prompt classification of this transient. SN 2018zd was later classified as a young Type IIn SN from a relatively high-quality spectrum obtained at LJT on March 05.74, 2018, which showed narrow emission lines superimposed on a blue continuum \citep{Zhang18}. 

Given the close distance and the young phase when the SN was discovered, we triggered a follow-up observational campaign of SN 2018zd with the LJT of YNAO, the Xing-Long 2.16\,m telescope (hereafter XLT), the Tsinghua-NAOC (National Astronomical Observatories of China) 0.8\,m telescope (hereafter TNT; \citealp{wang08,TNT}) at Xing-Long Observatory of the NAOC, the 0.6/0.9\,m Schmidt telescope (hereafter KOT) of the Konkoly Observatory of the Hungarian Academy of Sciences, and  the 0.6\,m Cassegrain reflector telescope (hereafter WT) at the Astronomical Observatory of the University of Wroclaw, Bialkow, Poland \citep{Mikolajczyk18}. The monitoring spanned from $t \approx 2$\,d to $t \approx 456$\,d (where $t$ denotes time after the explosion). Late-time spectra were also collected  with the 3.5\,m Astrophysical Research Consortium (hereafter ARC) telescope of the Apache Point Observatory (APO), and the Keck-I 10\,m telescope on Maunakea, Hawai'i. Moreover, the UV-optical photometry spanning from $t \approx 3$\,d to $t \approx +116$\,d obtained with UVOT on the Neil Gehrels {\it Swift} Observatory \citep{Gehrels04,Roming05} is also included here.

\begin{figure*}
\centering
\includegraphics[width=14cm,angle=0]{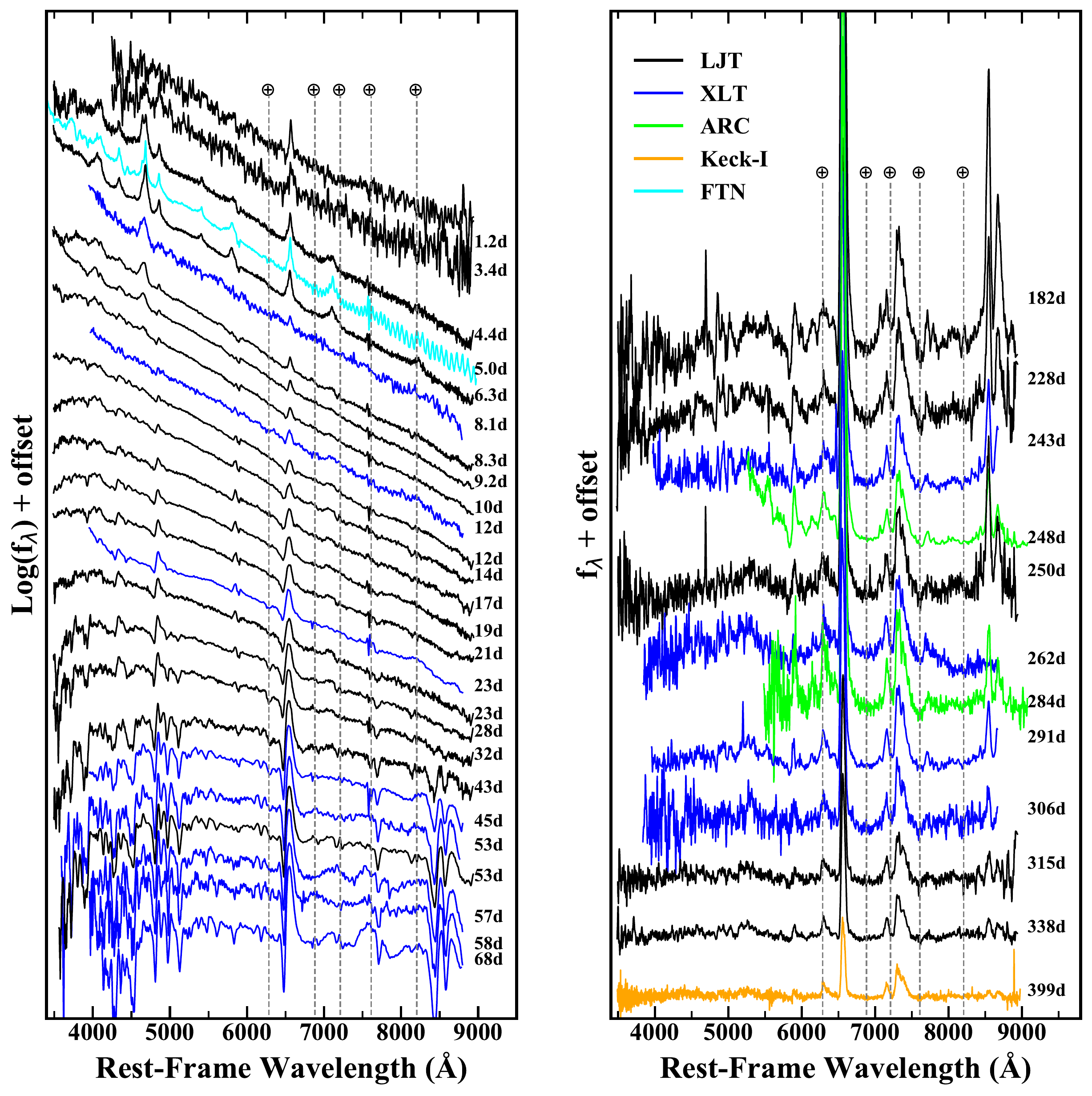}
\caption{Spectra of SN 2018zd obtained with LJT (black), XLT (blue), ARC (green), Keck-I (orange), and FTN (cyan), shifted vertically for clarity. Regions where telluric absorption was removed (sometimes leaving residuals) are marked with an Earth symbol. Various bin sizes were used for the different spectra, depending on the original signal-to-noise ratio. {\it Left:} early-phase and photospheric-phase spectra.
{\it Right:} nebular-phase spectra.}
\label{<spec>}
\end{figure*}

\subsection{Photometry}
\label{subsect:Photo}

SN 2018zd was densely observed by the LJT in the Johnson $UBV$ bands and the Sloan $gri$ bands, by the TNT and KOT in Johnson $BV$ and Cousins $RI$, and by the WT in Johnson $BV$ and Sloan $ri$. These images were reduced using IRAF\footnote{IRAF, the Image Reduction and Analysis Facility, is distributed by the National Optical Astronomy Observatory, which is operated by the Association of Universities for Research in Astronomy (AURA), Inc. under cooperative agreement with the National Science Foundation (NSF).} standard procedures, including corrections for bias, overscan, flat field, and removal of cosmic rays. The point-spread-function fitting method \citep{Stetson87} was used to measure the instrumental magnitudes of the SN.  We selected twelve local reference stars in the field of SN 2018zd, as marked in Fig. \ref{<img>}.  The instrumental magnitudes of these reference stars are converted to standard $UBVRI$ (Vega magnitudes) and $gri$ (AB magnitudes) through transformations established by observing a series of \citet{Landolt} and \citet{Smith02} standard stars on some photometric nights, respectively. These magnitudes are applied to calibrate the photometry of SN 2018zd, as presented in Table \ref{Tab:Pho_Ground}. The photometry of WT is calibrated via the Cambridge Photometric Calibration Server\footnote{http://gsaweb.ast.cam.ac.uk/followup}.

Photometry of SN 2018zd was obtained by the {\it Swift}/UVOT in three UV filters ($uvw2$, $uvm2$, and $uvw1$) and three broadband optical filters ($u$, $b$, and $v$) and is presented in Table \ref{Tab:Swiftpho}. These {\it Swift} images were reduced using the HEASoft\footnote{HEASoft, the High Energy Astrophysics Software http://www.swift.ac.uk/analysis/software.php} with the latest {\it Swift} calibration database\footnote{http://heasarc.gsfc.nasa.gov/docs/heasarc/caldb/swift/}.

All of the photometry is presented in Fig. \ref{<LC>}, which covers the first $\sim 450$\,d since discovery.  Note that the instrumental response curves of the UVOT optical filters do not follow exactly those of the Johnson $UBV$ system, especially in {\it Swift} $u$ where the UVOT coverage is wider than Johnson $U$.  Thus, we treat the $u$ and $U$ bands individually in the following calculation and analysis.

\begin{table*}
\small
\caption{Photometric Results for SN\,2018zd$^a$}
\scalebox{0.93}{
\begin{tabular}{cccccccccc}
\hline\hline
Band &$\lambda_{\rm eff}$ &FWHM (\AA)& $t_{\rm rise}$ (d) & $m_{\rm max}^b$\, (mag) & $M_{\rm max}$ (mag) & $M_{\rm end}$ (mag)  & $M_{\rm tail}$ (mag) & $\Delta m_{50}$ (mag)& $\gamma$ (mag\,[100\,d]$^{-1}$)\\
\hline
$uvw2$	&	1928	&	657	&	6.94(0.05)	& 12.71(0.03) &	-19.97(0.61) &  ...	&	...	&	5.94(0.05)	&	... \\
$uvm2$	&	2246	&	498	&	7.27(0.06)	&	12.83(0.05)	&	-20.06(0.67)	&	...	&	...	&	6.46(0.05)	&	...	\\
$uvw1$	&	2600	&	693	&	7.35(0.05)	&	12.81(0.03)	&	-19.61(0.54)	&	...	&	...	&	4.93(0.04)	&	...	\\
$u$	&	3465	&	785	&	8.27(0.05)	&	12.58(0.02)	&	-19.54(0.45)	&	...	&	...	&	3.30(0.03)	&	...	\\
$U$	&	3652	&	485	&	8.46(0.05)	&	12.97(0.02)	&	-19.14(0.44)	&	...	&	...	&	2.59(0.03)	&	...	\\
$B$	&	4312	&	831	&	9.13(0.05)	&	13.79(0.01)	&	-18.19(0.41)	&	-15.93(0.43)	&	-12.48(0.43)	&	1.53(0.03)	&	0.74(0.06)	\\
$g$	&	4754	&	1280	&	8.77(0.05)	&	13.62(0.01)	&	-18.29(0.39)	&	...	&	-13.02(0.45)	&	1.24(0.03)	&	0.75(0.07)	\\
$V$	&	5341	&	827	&	8.98(0.05)	&	13.55(0.01)	&	-18.26(0.36)	&	-16.82(0.38)	&	-13.43(0.38)	&	0.93(0.03)	&	0.90(0.07)	\\
$r$	&	6196	&	1150	&	9.21(0.05)	&	13.59(0.01)	&	-18.15(0.34)	&	-17.14(0.36)	&	-14.49(0.38)	&	0.63(0.03)	&	1.09(0.06)	\\
$R$	&	6315	&	1742	&	9.28(0.07)	&	13.45(0.01)	&	-18.28(0.33)	&	-17.31(0.36)	&	-14.37(0.36)	&	0.60(0.02)	&	0.80(0.06)	\\
$i$	&	7690	&	1230	&	9.67(0.05)	&	13.74(0.01)	&	-17.88(0.30)	&	-17.22(0.32)	&	-14.06(0.35)	&	0.37(0.02)	&	0.96(0.06)	\\
$I$	&	8752	&	1970	&	10.18(0.08)	&	13.36(0.01)	&	-18.19(0.28)	&	-17.50(0.31)	&	-14.66(0.33)	&	0.35(0.02)	&	1.12(0.09)	\\
\hline
\hline
\label{Tab:LC_res}
\end{tabular}
}\\
$^a${The $uvw2$, $uvm2$, $uvw1$, and $u$ bands of {\it Swift}/UOVT, and the Johnson $UBVRI$ bands, are in the Vega magnitude system. SDSS $gri$ bands are in the AB magnitude system. The numbers given in brackets are the 1$\sigma$ uncertainties.}
$^b${Derived by fitting a cubic polynomial to the points around maximum brightness.}
\end{table*} 

\subsection{Spectroscopy}
\label{subsect:Sp}
Fig. \ref{<spec>} shows the spectral sequence of SN 2018zd spanning from $t\approx+2$\,d to $t\approx+400$\,d. The journal of observations is given in Table \ref{Tab:Spec_log}, including twenty-two spectra from LJT (with the YFOSC), twelve from XLT (with the Beijing Faint Object Spectrograph and Camera, BFOSC), two from ARC (with the Dual Imaging Spectrograph, DIS), and one from Keck-I (with the Low-Resolution Imaging Spectrometer, LRIS; \citealt{oke95}). All of these spectra are calibrated in both wavelength and flux, and they are corrected for telluric absorption and redshift. Spectra were obtained with a slit oriented along the parallactic angle to minimise the effects of atmospheric dispersion \citep{Filippenko82}; moreover, the continuum shape was further corrected with multiband photometry. One spectrum (presented at the transient name system, TNS\footnote{ https://wis-tns.weizmann.ac.il/object/2018zd}), obtained using the FLOYDS spectrograph on the Las Cumbres Observatory 2\,m Faulkes Telescope North (FTN) on Haleakala, Hawai'i, is also plotted in this figure to help elucidate the rapid evolution at very early times.

\subsection{Distance and extinction}
\label{sect:DandE}

The distance $D$ of SN 2018zd adopted in the following calculation is $18.4 \pm 4.5$\,Mpc, derived from the averaged measurements of the host galaxy (NGC 2146) listed in the NASA/IPAC Extragalactic Database (NED\footnote{http://ned.ipac.caltech.edu}). For example, the average estimation from the Tully-Fisher relation is $D = 19.2 \pm 8.8$\,Mpc \citep{Bottinelli84,Giraud85,Tully88,Schoniger94,Tutui97}, the measurement via the observed angular radius of typical globular clusters in NGC 2146 (comparing with the physical radii of Milky Way globulars) is $18.0 \pm 1.8$\,Mpc \citep{Adamo12}, and the result from the radial velocity of NGC 2146 ($v = 1219 \pm 16$\,km\,s$^{-1}$, after correcting for Local Group infall into the Virgo Cluster, the Great Attractor, and the Shapley Supercluster \citealp{Mould00}) is $17.98 \pm 1.26$\,Mpc with H$_0 = 67.8$\,km\,s$^{-1}$\,Mpc$^{-1}$ \citep{Planck14}.  

In our analysis, we use the interstellar \ion{Na}{i}~D absorption to estimate the reddening toward SN 2018zd. Two \ion{Na}{i}~D absorption systems are detected in the spectra of SN 2018zd, with similar equivalent width (EW), $\sim 1.3\pm0.1$\,\AA\ (see Fig. \ref{<Sp_comp>}), suggesting significant extinction due to the host galaxy and the Milky Way.  The minimum reddening value derived from the existing  empirical correlations between reddening and EW of \ion{Na}{i}~D, $E(B-V) = 0.16\,{\rm EW} - 0.01$\,mag \citep{Tura03}, is $E(B-V)_{\rm host} = E(B-V)_{\rm MW} = 0.198$\,mag. However, the Galactic reddening derived by \citet{Schlafly11} is only  $E(B-V)_{\rm MW} = 0.085$\,mag.   Considering the large scatter of the reddening measurement via \ion{Na}{i}~D absorption (e.g., \citealp{Poznanski11}), we assume $E(B-V)_{\rm host} = E(B-V)_{\rm MW}$ because of the similar EW of \ion{Na}{i}~D from the host galaxy and the Milky Way. Thus, a total  reddening of  $E(B-V) = 0.17 \pm 0.05$\,mag with the extinction law $R_{V} = 3.1$ is adopted in this paper.  This matches the estimate via the $V-I$ colour in Section \ref{subsect:CC}.  We caution, however, that the luminosity of SN 2018zd may be underestimated because of the conservative host-galaxy reddening adopted in our analysis.

\section{Analysis of the Photometry}
\label{sect:LC}

\subsection{Explosion date and a shock-breakout signal}
\label{subsect:rise} 
\begin{figure}
\centering
\includegraphics[width=8cm,angle=0]{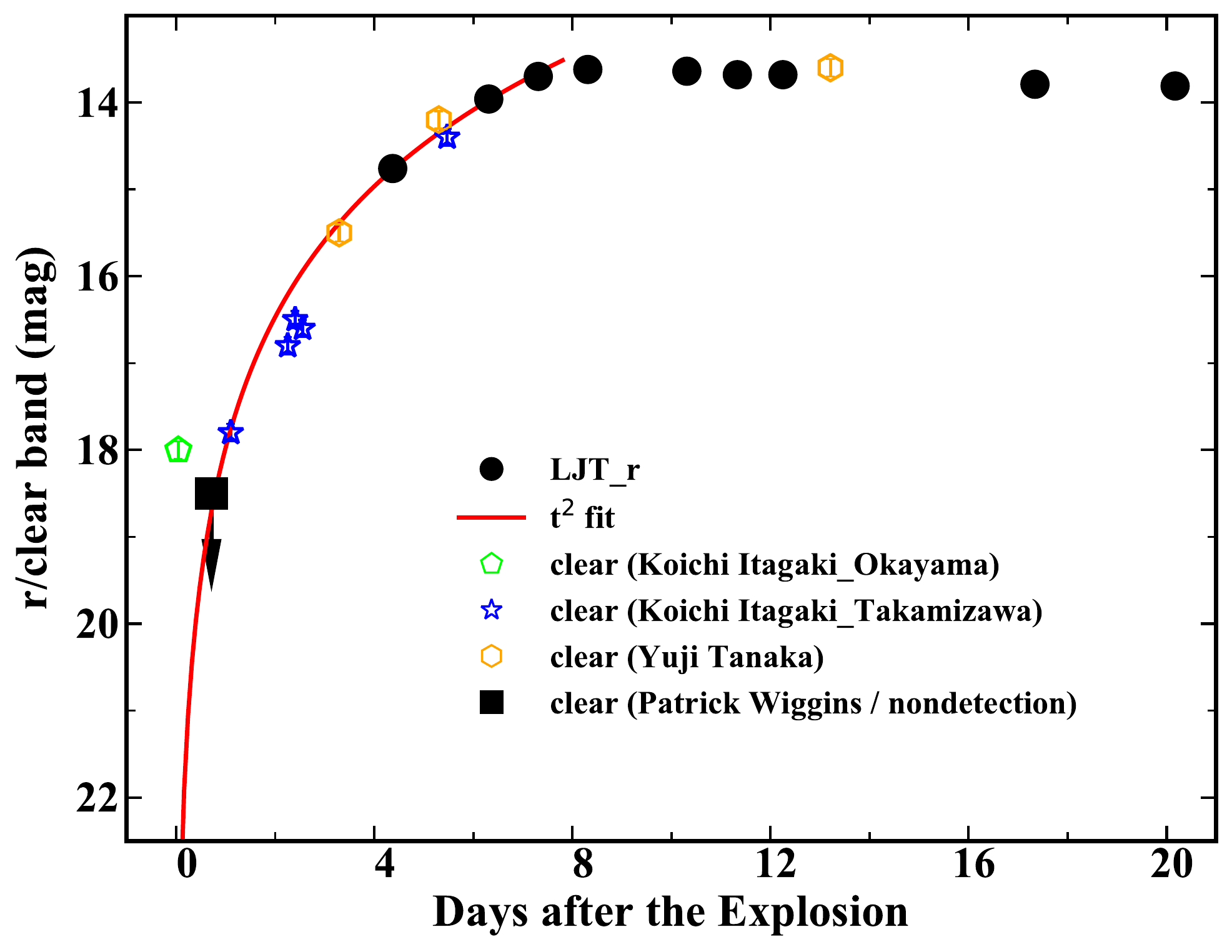}
 \caption{Expanding fireball fit of the early detections of SN 2018zd in the clear band and the $r$ band, where the first detection by Itagaki cannot be fitted.}  
\label{<t2>}
\end{figure}

To better estimate the explosion date of SN 2018zd,  we also collect the clear-band data obtained by amateur astronomers in the early phase of detections\footnote{``Bright Supernovae,'' www.rochesterastronomy.org/sn2018/sn2018zd.html}, as listed in Table \ref{Tab:LC_ama}. The clear-band photometry roughly matches  that of the $r$ band of LJT, as shown in Fig. \ref{<t2>}.  

A simple expanding fireball model, $F(t)=F_1 \times (t - t_0)^2$, is applied to fit observations at $t \lesssim 8$\,d except for the earliest detection. The explosion date derived by the  expanding  fireball model is MJD = 58178.39$^{+0.15}_{-0.50}$, which implies that SN 2018zd was first detected (by Itagaki; see below) only 3.6\,hr after the explosion. It is reasonable that SN 2018zd was very young at the first detection because its brightness increased by $\sim 5$\,mag when approaching the peak.  

Note that Itagaki had a prediscovery detection (obtained with a 0.35\,m telescope at Okayama Observatory) at the site of SN 2018zd roughly one day before his official discovery image, with an unfiltered magnitude of 18.0. About 13\,hr after this earliest detection, however, Patrick Wiggins\footnote{https://wis-tns.weizmann.ac.il/object/2018zd\#comment-wrapper-1927} reported a nondetection of SN 2018zd in an unfiltered image with an upper limit of 18.5\,mag. Such an intraday dip is very likely related to the shock-breakout phenomenon (as seen in KSN 2011d; \citealp{Garnavich16}). 

\begin{figure*}
\centering
\includegraphics[width=16cm,angle=0]{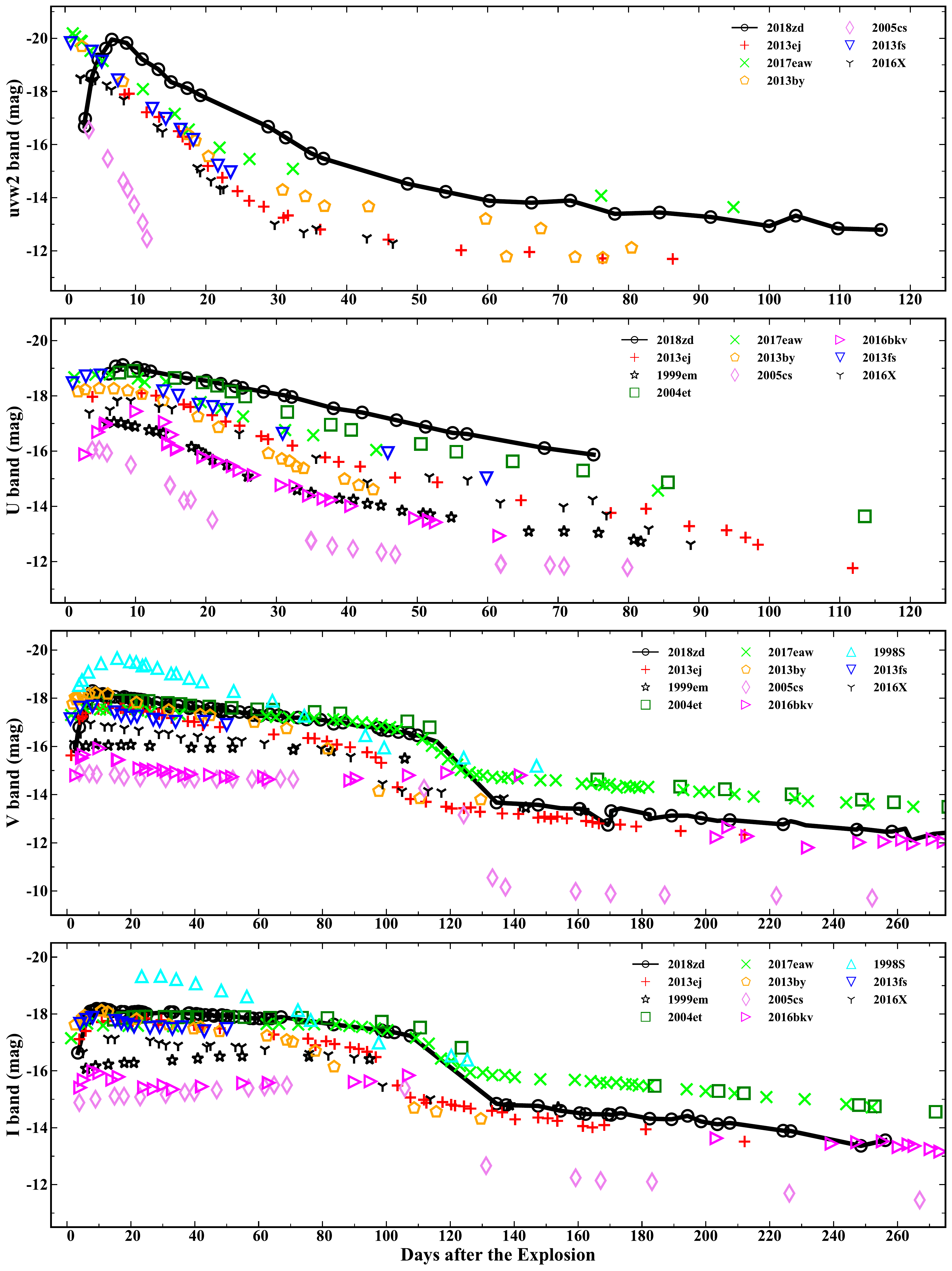}
 \caption{Light-curve comparisons (in absolute magnitude) between SN 2018zd with some well-studied SNe~II, including the standard SNe~IIP 1999em \citep{Hamuy01,Leonard02,Elmhamdi03} and 2004et \citep{Sahu06,Misra07}, the normal SN~IIP 2017eaw (\citealp{Rui19,Szalai19}, and unpublished data collected with LJT and TNT), the fast-declining SNe~II 2013ej \citep{Huang15,Dhungana16,Yuan16} and 2013by \citep{Valenti15}, the low-velocity and low-luminosity SN~IIP 2005cs \citep{Pastorello06,Pastorello09}, and SNe~II showing interaction signatures in their spectra such as SNe~2013fs \citep{Yaron17,Bullivant18}, 2016bkv \citep{Hosseinzadeh18,Nakaoka18}, and 1998S \citep{Leonard00,Fassia00,Poon11}.}
\label{<LCabs>}
\end{figure*}

\begin{figure}
\centering
\includegraphics[width=8cm,angle=0]{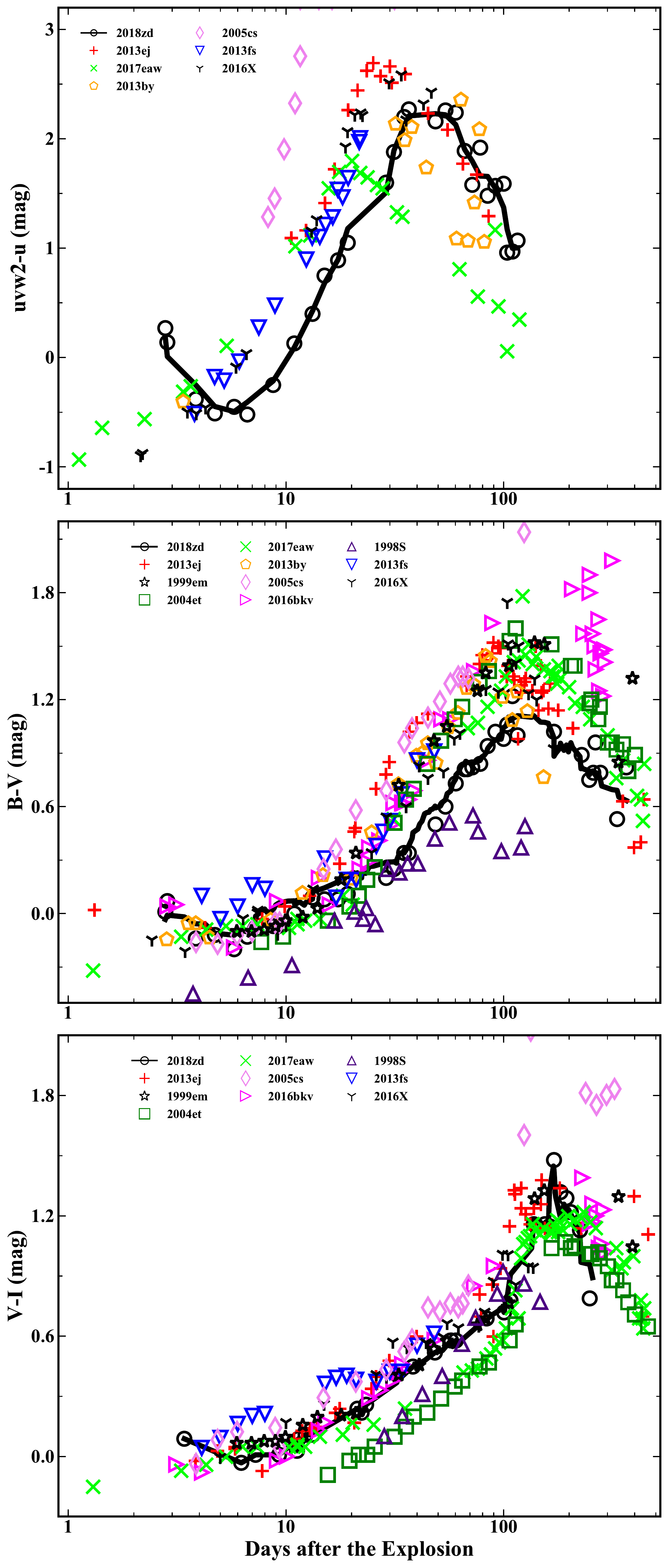}
 \caption{The $uvw1-u$, $B-V$, and $V-I$ colours of SN 2018zd compared with those of the sample presented in Fig. \ref{<LCabs>}. All of these colour curves have been corrected for host-galaxy and Milky-Way extinction. }
\label{<CC>}
\end{figure}

\subsection{Photometric results}
\label{sect:Phores}
Table \ref{Tab:LC_res} lists estimates of the explosion time to peak brightness ($t_{\rm rise}$), the apparent ($m_{\rm max}$) and absolute ($M_{\rm max}$) magnitude at maximum brightness, the absolute magnitude at the end of the plateau phase ($M_{\rm end}$, at $t = 125 $\,d, in the case of SN 2018zd) and at the beginning of the radiative tail ($M_{\rm tail}$, at $t = 155$\,d), the decline rate from peak to 50\,d later ($\Delta m_{50}$), and the  decline rate per 100 d ($\gamma$) during the radioactive tail (at $t > 150$\,d, when the light curve is powered by radioactive decay and is governed by $\gamma$-ray trapping) of SN 2018zd in all bands presented in Fig. \ref{<LC>}.

Compared with the sample from  \cite{Pritchard14}, the absolute peak magnitudes of SN 2018zd are located at the luminous end of SNe~IIP and SNe~IIL, and the faint end of SNe~IIn. The UV peak of SN 2018zd ($M_{\rm UV} \approx -20$\,mag) is brighter than that of SNe~IIP ($M_{\rm UV} \approx -18$\,mag), and is close to the average brightness of SNe~IIn ($M_{\rm UV} \approx -20$\,mag). The optical brightness of SN 2018zd (e.g., $M_{\rm Op} \approx -18$\,mag, in the $B$ and $V$ bands) lies between the bright end of SNe~IIP ($-18 < M_{\rm Op} < -15$\,mag) and the faint end of SNe~IIn ($-20 < M_{\rm Op} < -18$\,mag; \citealp{Pritchard14}). \cite{Richardson02} found that the average $B$-band peak absolute magnitude is $M_B = -17.0 \pm 1.1$\,mag for 29 SNe~IIP and $M_B = -18.0 \pm 0.9$\,mag for 19 SNe~IIL. This indicates that SN 2018zd is brighter than most SNe~IIP and comparable to SNe~IIL in the optical. Furthermore, considering SNe~IIP and IIL as a continuous family, SN 2018zd is at the bright end of the distribution ($-13.77<M_V<-18.29$\,mag; \citealp{Anderson14}).  

As shown in Fig. \ref{<LC>}, the light curve of SN 2018zd had a fast decline from the peak to the plateau phase.  \cite{Faran14b} suggested that SNe~II with $\Delta m_{50} > 0.5$\,mag in the $V$ band can be classified as SNe~IIL. Based on such a criterion, SN 2018zd could be placed into the SN~IIL group because of a relatively quick decline (i.e., $\Delta m_{50}^V = 0.93\pm0.03$\,mag). However, a further significant flux drop can be found at $t\approx115$\,d in the optical light curves, with a decline of $\sim 3$\,mag when entering the radioactive-decay tail. Such a significant flux drop has been regarded as a typical feature of SNe~IIP when hydrogen recombination process ends in the envelope. 

Nevertheless, \cite{Anderson14} and \cite{Valenti15} proposed that if all SNe~IIL were monitored sufficiently long, they would exhibit a significant drop in their late-time light curves. SN 2018zd is an excellent example supporting the above argument because of the SN~IIL-like initial decline rate and the SN~IIP-like drop from the end of plateau to the radioactive tail.

We notice that the UV light curves of SN 2018zd showed an unusually slow increase to peak compared to the optical, reaching maximum brightness $\sim 1$ week after explosion. However, SNe~II usually reach their UV maximum much earlier (at $t \approx 1$--3\,d) than in the optical ($\sim 5$\,d in  $U$ and $\sim 10$\,d in $I$). For example, \cite{Pritchard14} noted that most of their sample were observed before the $V$-band maximum, but only a few before UV maximum even though some were monitored at quite young phases.

\subsection{Morphology of the light curves}
\label{subsect:PP}
The  multiband light curves of SN 2018zd are shown  in Fig.  \ref{<LCabs>}, together with those of some well-observed SNe~II. The morphology comparison  confirms the results derived  in Section \ref{sect:Phores}:\\
(1) The UV emission  of SN 2018zd rises slowly; it reaches the $uvw2$-band peak $\sim3$--6\,d later than the other comparison SNe. The rise time of these  SNe in the UV band is uncertain because their peaks occur near the time of the first detection. Thus, these events may have reached their maximum brightness a few days earlier than shown  in this figure. \\
(2) The peak brightness of SN 2018zd is close to that of SNe~IIP at the bright end, and it is close to that of SNe~IIL and SNe~IIn. Among the comparison SNe~II, only SN~IIn 1998S shows a more luminous peak than SN 2018zd.  \\
(3) SN 2018zd has an intermediate decline rate during the plateau phase, faster than that of regular SNe~IIP (e.g., SN 1999em, SN 2004et, and SN 2017eaw), but slower than that of fast-declining SNe~II (e.g., SN 2013ej and SN 2013by).\\
(4) SN 2018zd exhibits the most significant drop from the peak to the end of the plateau phase, even greater than that of SN 2013ej and SN 2013by.

\begin{figure*}
\centering
\includegraphics[width=17cm,angle=0]{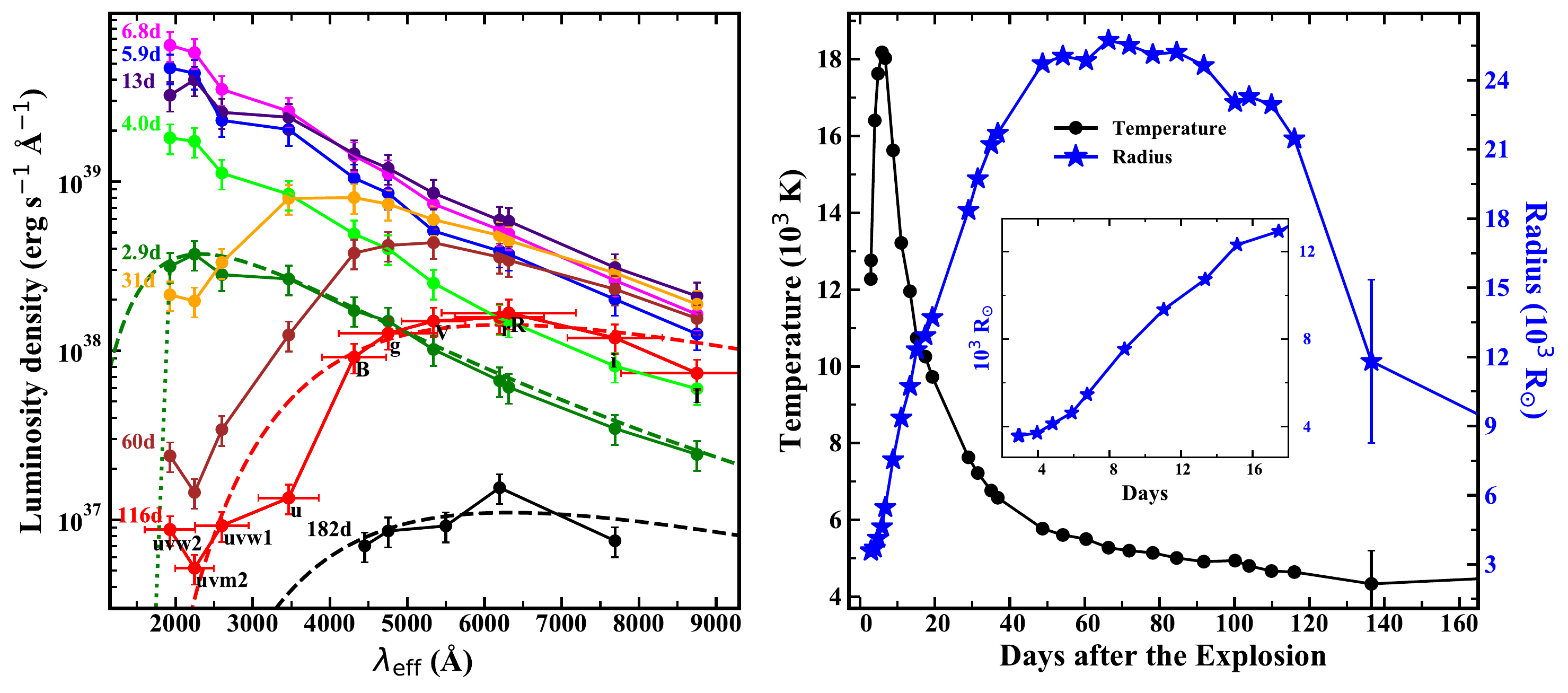}
 \caption{{\it Left:} SED of SN 2018zd at selected phases. The dashed lines are examples of blackbody fits to the observations at $t = 2.9$\,d and 182\,d.  The dotted-line is the flux extrapolation of the flux at the wavelength shorter than 1928\AA, the flux at the wavelength shorter than 1216\AA\,  goes to zero due to the absorption of H-rich atmosphere. The uncertainty of each SED is about 20\%, including errors in photometry, extinction, and distance. {\it Right:} Temporal evolution of the temperature and the photospheric radius of SN 2018zd. The inset shows a close-up view of the radius at $t<18$\,d. The typical uncertainty of the temperature is about 20\%, including errors in photometry, extinction, and blackbody fitting. The typical uncertainty of the radius is about 30\%, including errors in temperature and distance, see the sample errors at $t \approx 137$ d.  }
\label{<SED>}
\end{figure*}

\begin{figure*}
\centering
\includegraphics[width=17cm,angle=0]{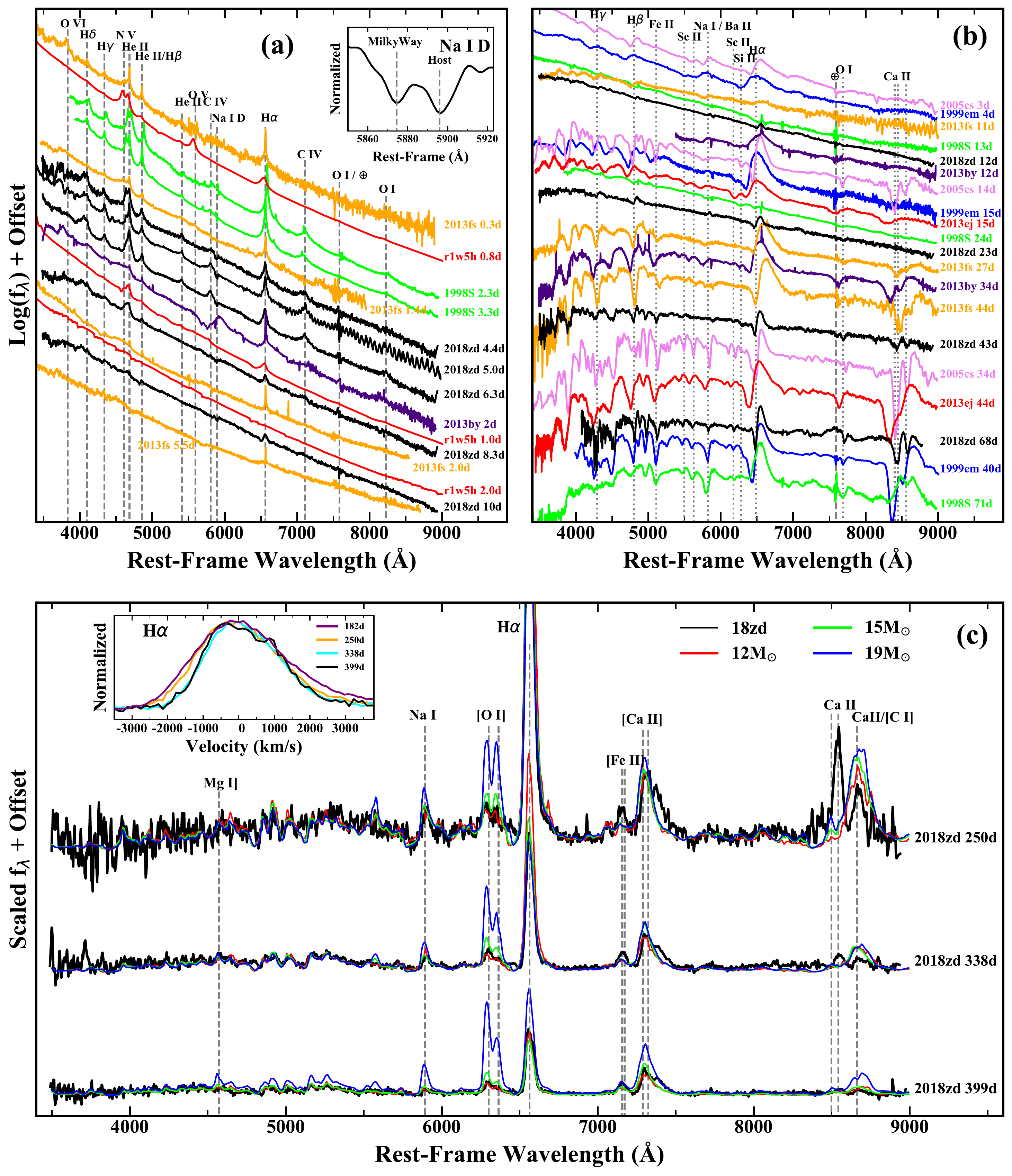}
 \caption{Spectral comparison at three phases. (a) Flash-ionisation-phase spectra of SN 2018zd compared with spectra of SN 1998S \citep{Leonard00,Fassia01,Shivvers15},  SN 2013by \citep{Valenti15}, SN 2013fs \citep{Yaron17}, and model spectra r1w5h of \citep{Dessart17}. The inset shows a close-up of the region of Na~I~D absorption in the spectrum of SN 2018zd at $t\approx5$\,d. (b) Photospheric-phase spectra of SN 2018zd along with those of SN 1998S, SN 1999em\citep{Hamuy01,Leonard02}, SN 2005cs \citep{Pastorello06}, SN 2013by , SN 2013ej \citep{Dhungana16}, and 2013 SN 2013fs. (c) Three nebular-phase spectra of SN 2018zd (at 250\,d, 338\,d, and 399\,d) compared with model spectra corresponding to different progenitor masses (\citealt{Jerkstrand12}; 12\,M$_{\odot}$ at 250\,d, 306\,d, and 400\,d; 15\,M$_{\odot}$ at 250\,d, 332\,d, and 400\,d; and 19\,M$_{\odot}$ at 250\,d, 332\,d, and 369\,d). The model spectra are scaled to the distance of SN 2018zd and its $^{56}$Ni mass. The inset shows the velocity distribution of the H$\alpha$ emission of SN 2018zd at $t = 182$\,d, 250\,d, 338\,d, and 399\,d; the instrumental resolution was removed via FWHM$_{\rm corr}$ = (FWHM$^2_{\rm obs}$ $-$ FWHM$^2_{\rm inst})^{1/2}$, where the instrumental FWHM was measured from the night-sky emission lines. In panels (a) and (c), dashed lines mark spectral features at rest; in panel (b), dotted lines indicate spectral features at $v = -3500$\,km\,s$^{-1}$. In panels (a) and (b), residuals from incomplete removal of telluric absorption are marked with an Earth symbol. The host-galaxy redshift and the  extinction from the host galaxy and the Milky Way have been removed.}
\label{<Sp_comp>}
\end{figure*}

\subsection{Colour curves}
\label{subsect:CC}
SN 2018zd shows bluer $uvw2-u$ and $B-V$ colours at $t \lesssim 50$\,d than normal SNe~IIP, as presented in Fig. \ref{<CC>}, which might result from the contribution of CSM interaction. Based on a similar argument, it is not surprising that the colour of SN 1998S is bluer than that of SN 2018zd because the former might have stronger interaction.
 
Different comparison samples of SNe II generally exhibit similar colour evolution but with some notable differences. The most obvious is that SN 2018zd shows a very clear reversal (``U-turn'') in the $uvw2-u$ colour curve at early times. A similar but weaker feature is present in the $B-V$ and $V-I$ colour curves. This U-turn behaviour indicates that the temperature increase of SN 2018zd is consistent with the temperature evolution derived in the following section. Besides, the decreased extinction due to dust destruction at the early phase of the explosion could also explain this U-turn behavior.  However, no evidence of dust destruction (e.g., the variation of   \ion{Na}{i}  D absorption from the host galaxy) is found in our low-resolution spectra.

The $uvw2-u$ colour U-turn is not typical of the SN~II family even among those with very early-time observations, such as SN 2017eaw and SN 2016X. It is suspected that SN 2013ej, SN 2013by, and SN 2016bkv might also have similar $uvw2 - u$ colour evolution, as they show the $U$-turn feature in the $B-V$ colour curve; however, their UV data are either absent or too poorly sampled for a detailed comparison.

 The $uvw2 - u$ colour of SN 2018zd becomes bluer again at $t > 40$\,d, a reverse ``U-turn'' shape that is due to the increased H Balmer absorption within the $u$ band and the decreased metal line-blanking in the $uvw2$ band (e.g., see the UV-optical spectral evolution in Figure 2 of \citealp{Dessart13}).  The similar reverse ``U-turn''  seen in the $B-V$ and $V-I$ colours at $t \gtrsim 150$ d is related to the evolution of emission lines in the nebular phase.  These reverse ``U-turn'' shapes seen in the colour curves do not imply that the temperature is rising. 

The intrinsic colour of SNe~II could be another way to estimate extinction. For example, \citet{Hamuy03} suggested using the $V-I$ colour as a better reddening indicator than the $B-V$ colour because it is expected to be less sensitive to metallicity effects. The  $V-I$ colour of SN 2018zd roughly follows that of SN 1999em, SN 2013ej, SN 2016X, and SN 2017eaw at $10 < t < 100$\,d. This suggests a smaller effect of interaction in the $V-I$ colour of SN 2018zd than in the bluer colour. Thus, we could estimate the extinction via an empirical relation, such as $A_V(V-I) = \beta_V[(V-I) - 0.656]$ \citep{Olivares10}, where the $V-I$ colour corresponds to the colour of a given SN 30\,d before the end of the plateau, and $\beta_V = A_V/E(V-I)$ is related to $R_V$ (e.g., $\beta_V = 2.518$ for $R_V=3.1$). Based on the $V-I$ colour of SN 2018zd measured at the appropriate phase ($t \approx 65$\,d), the total extinction is estimated to be $A_V^{\rm total} = 0.52 \pm 0.10$\,mag, and $E(B-V)_{\rm total} = 0.168 \pm 0.033$ for $R_V = 3.1$. These values are consistent with the reddening adopted in this paper.

\subsection{Spectral energy distribution}
\label{subsect:SED}

The left panel of Fig. \ref{<SED>} displays the spectral energy distribution (SED) of SN 2018zd at some selected epochs.  At $t \lesssim 116$\,d, when {\it Swift}-UVOT observations are available, we construct the SED from the observed fluxes in the UV through optical passbands (covering the wavelength from $\sim 1600$ to $\sim 9700$\,\AA) obtained at the same phase. The missing data are obtained through interpolation of the adjacent data.  The SEDs at $t>155$\,d are constructed only from the $BVRI$ or the $BgVri$ photometry. 

The SED peaks toward shorter wavelengths in the first few days and then toward longer wavelengths, which conforms to an initial heating followed by cooling processes as indicated in the temperature curve  presented in the right panel of Fig. \ref{<SED>}. The temperature of SN 2018zd is derived via  blackbody fitting of the SED; it increases from $\sim 12,000$\,K to $\sim 18,000$\,K in the first week after the explosion. This behaviour might be related to CSM interaction at early times, as is also indicated by the early spectra (see Fig. \ref{<Sp_comp>}).  At these phases, the forward shock accelerates and heats the shocked wind/CSM and the reverse shock reheats the outer layers of the ejecta \citep{Branch17}.

For SNe II, it is not common to see the temperature increase at early times.  \cite{Hosseinzadeh18} reported the unexpected temperature rise of SN 2016bkv during a similar period.  However, their conclusion is uncertain, owing to a lack of near-UV photometry. Thus, SN 2018zd could be the first SN~II for which there is conclusive evidence of a rising temperature at early phases.

The right-hand panel of Fig. \ref{<SED>} also displays the photospheric radius ($R$) of SN 2018zd calculated from the relation $L = 4 \pi \sigma R^2 T^4$, where the luminosity $L$ is derived by integrating the SED from 1600\,\AA\ to 27,000\,\AA; flux outside the wavelength range of photometric coverage is extrapolated based on a blackbody spectrum.  

The photospheric radius of SN 2018zd is $R = (3.57\pm0.50) \times 10^3\,{\rm R}_{\odot}$ at $t \approx 2.9$\,d, increasing slowly during the temperature rising phase.  After $\sim 40$\,d of faster expansion, SN 2018zd reached maximum photospheric radius ($R \approx 2.5 \times 10^4\, {\rm R}_{\odot}$) at $t \approx 45$\,d and remained almost unchanged during the next two months; thus, recombination provided the dominant source of energy at this phase. The recombination temperature is $\sim 6000$\,K, so the ejecta consist of a mixture of hydrogen and helium \citep{Grassberg76}. The photospheric radius decreases when recombination ends and the SN envelope becomes transparent.

\section{Spectroscopic Analysis}
\label{sect:SP}

Fig. \ref{<Sp_comp>} displays the spectral comparison of SN 2018zd with a selected sample and model spectra obtained at different phases, including the flash-ionisation, photospheric, and nebular phases.

\subsection{Flash-ionisation phase}
\label{subsect:spinter}
Panel (a) of Fig. \ref{<Sp_comp>} displays early-time spectra of SN 2018zd at $t \lesssim 10$\,d, together with that of SN 1998S, SN 2013fs, and the r1w5h model spectra given by \cite{Dessart17} at similar phases. The spectra of SN 2018zd are dominated by flashed-ionised features \citep{Gal-Yam14} such as the narrow emission lines of hydrogen, \ion{N}{v} $\lambda\lambda4334$, 4641, \ion{He}{ii} $\lambda4686$, \ion{He}{ii} $\lambda4860$, \ion{C}{iv} $\lambda\lambda5801$, 5812, and \ion{C}{iv} $\lambda7110$. These features are generated by the surrounding wind material which was ionised by X-rays from the shocked ejecta \citep{Branch17}. The first two low-quality spectra of SN 2018zd in Fig. \ref{<spec>} show possible H$\alpha$ emission, suggesting that the interaction starts $\sim 1.2$\,d after explosion. 

The flash features in SN 2018zd appear to evolve more slowly than in SN 2013fs. A noticeable change from $t\approx4.4$\,d to $t\approx6.3$\,d is the weakening of \ion{N}{v} $\lambda\lambda4634$, 4641, which suggests a decreasing ionisation. However, this seems to conflict with the increasing surface temperature inferred at this phase (see  Section \ref{subsect:SED}); perhaps the temperature derived from the SED may not fully reflect radiation in the ionisation region.  

The spectra of SN 1998S at $t\approx2.3$\,d and 3.3\,d are similar to that of SN 2018zd at $t\approx4.4$\,d. The similarity of the two spectra of SN 1998S (separated by 1\,d) might suggest slower spectral evolution than that of SN 2018zd. In contrast, SN 2013fs went through very rapid changes during the first few days after the explosion. The narrow emissions seen in SN 2013fs showed pronounced variations on a timescale of hours and almost vanished $\sim 2$\,d after the explosion.

\cite{Dessart17} applied  radiation hydrodynamics and nonlocal-thermodynamic-equilibrium (NLTE) radiative transfer to produce observational features of SNe~II with short-lived flash spectra considering different physical states of the progenitor before the explosion. The ionisation features produced in the r1w5h (where r1 means the progenitor star radius with $R_{\star}=501\,{\rm R}_{\odot}$, w5 means a wind mass-loss rate of $\dot{M}=5\times10^{-3}\,{\rm M}_{\odot}\,{\rm yr}^{-1}$, and h means the atmospheric density scale height of $H_{\rho}=0.1\,R_{\star}$) model is the closest to that seen in SN 2018zd among their models.

One can see that the  r1w5h model can reproduce the dominant spectral features observed in SN 2018zd, but the model evolves too fast. For example, the model spectrum at $t\approx0.8$\,d shows a similar continuum and spectral features (e.g., \ion{N}{v} and \ion{He}{ii} lines) as the  $t\approx4.4$\,d spectrum of SN 2018zd. The Balmer lines in the model spectrum are weaker than those  of SN 2018zd. The model spectrum at $t=1.0$\,d exhibits line features similar to those in the $t \approx 8.3$\,d spectrum of SN 2018zd. The continuum of the featureless model spectrum at $t=2.0$\,d has a slope similar to that of SN 2018zd at $t \approx 8.3$\,d.

The slower spectral evolution and the longer-lived flash features of SN 2018zd in comparison with SN 2013fs and the r1w5h model are consistent with the slower light-curve rise seen in SN 2018zd. This difference might  suggest a more massive and  extended stellar wind surrounding SN 2018zd. On the other hand, the spectral evolution of SN 2018zd during the flash phase seems to be  faster than that of SN 1998S. Therefore, we propose that SN 2018zd might have an environment (e.g., the mass of wind/CSM) lying between that of SN 2013fs and SN 1998S in mass immediately before the explosion.

\subsection{Photospheric phase}
\label{subsect:spps}
Panel (b) of Fig. \ref{<Sp_comp>} displays spectra of SN 2018zd from $t\approx 12$\,d to $t\approx 68$\,d compared with photospheric-phase spectra of SNe 1999em and 2005cs, as well as with spectra of SNe 1998S, 2013by, 2013ej, and 2013fs at similar epochs. 

At this phase, spectra of SN 2018zd are still characterised by a featureless blue continuum, likely due to continuous heating by CSM interaction. Consequently, the spectrum of SN 2018zd at $t \approx 12$\,d looks younger than that of SN 2005cs at $t\approx 3$\,d and SN 1999em at $t \approx 4$\,d; it has a bluer colour and a weaker P-Cygni H$\alpha$ profile. Given its blue and featureless spectrum, SN 2018zd might be classified as a very young SN~II (e.g., 1--2\,d after the explosion) if it had not been spectrally observed until $t \approx 12$\,d.  Moreover, cross-correlation with a library of SN spectra using the supernova identification code (SNID; \citealp{Blondin07}) shows that spectra of SN 2018zd at $t \approx 23$, 43, and 68\,d respectively match those of SN 2005cs at $t \approx 4$, 10, and 13\,d, SN 1999em at $t \approx 6$, 11, and 33\,d, and SN 2004et at $t \approx 7$, 9, and 40\,d.

\cite{Khazov16} reported that 14\% of their SN~II sample observed at $t<10$\,d show flash spectral features. This fraction might be underestimated because  interaction makes SNe~II look somewhat younger than their real age. For example, SN 2018zd might have been counted as a young SN~II without flash features if both the discovery date and classification date had been delayed by $\sim 10$\,d.

At $t \approx 13$\,d, the spectrum of SN 2018zd looks similar to that of SN 1998S at a comparable phase, but it shows a more pronounced P-Cygni profile of the H$\alpha$ line than the latter. The spectrum of SN 2013fs at $t \approx 13$\,d exhibits a redder continuum and more noticeable Balmer lines than that of SN 2018zd, suggesting short-lived contribution of ionisation/interaction in the former, as seen in the flash phase. 

The evolution in the light curve, surface radius, and temperature indicate that SN 2018zd starts recombination at $t \approx 40$\,d. After this phase, the spectrum of SN 2018zd (e.g., $t \approx 43$\,d) evolves to be like that of SNe~IIP. For example, the spectrum of SN 2018zd at $t \approx 68$\,d is somewhat similar to that of SN 1999em at $t\approx 40$\,d, but with narrower and shallower spectral features. The contribution of CSM interaction still exists $\sim 2$\,months after the explosion, as suggested by the bluer $uvw2-u$ and $B-V$ colours of SN 2018zd at the same phase.

The EW ratio between the blue absorption wing and red emission wing of H$\alpha$ in spectra of SN 2018zd ($R_{\rm H\alpha} \approx 0.3$) is smaller than that of SN 1999em ($R_{\rm H\alpha} \approx 0.6$) but larger than that of SN 2013by ($R_{\rm H\alpha} \approx 0.1$), at $t \approx 20$ -- 30\,d. SN 1998S does not show a well-developed H$\alpha$ absorption component at $t\approx71$\,d ($R_{\rm H\alpha}<0.1$). The relatively weak H$\alpha$ absorption of SN 2013by and SN 19998S is usually seen in other SNe~IIL \citep{Schlegel96}. The narrower H$\alpha$ profile and the lower $R_{\rm H\alpha}$ of SN 2018zd might be related to a slower bulk expansion velocity than that of  typical SNe~IIP \citep{Dessart09}. It is consistent with the low expansion velocity of SN 2018zd seen in Section \ref{subsect:vel} and \ref{sect:Diver}.

\subsection{Nebular phase}
\label{subsect:spnb}
In the nebular phase, spectra of SN 2018zd evolve to be more like that of a normal SN~IIP, dominated by emission lines of [\ion{O}{i}], H$\alpha$, [\ion{Fe}{ii}], [\ion{Ca}{ii}], and \ion{Ca}{ii}. Panel (c) of Fig. \ref{<Sp_comp>} shows three nebular spectra of SN 2018zd and model spectra of different progenitor masses \citep{Jerkstrand12}. Comparison of the strength of [\ion{O}{i}] $\lambda\lambda$6300, 6364 between the observation and model suggests that the progenitor mass of SN 2018zd is in the range from 12\,M$_{\odot}$ to 15\,M$_{\odot}$, but preferring a lower mass.

The inset panel shows a close-up view of the H$\alpha$ emission; the profile of has been corrected for the instrumental broadening effect, FWHM$_{\rm corr}$ = (FWHM$^2_{\rm obs}$ $-$ FWHM$^2_{\rm inst})^{1/2}$, where FWHM$_{\rm obs}$ is the observed full width at half-maximum intensity (FWHM), and FWHM$_{\rm inst}$ is the instrumental FWHM. This emission becames progressively narrower from $t\approx182$\,d to $t\approx338$\,d, and then it levels off. The asymmetric H$\alpha$ emission at $t\approx399$\,d might imply an asymmetric structure in the inner part of this SN.

Maguire et al. (2012) reported a relation that can be  roughly used to  estimate the $^{56}$Ni mass from FWHM$_{\rm corr}$: $M(^{56}{\rm Ni}) = A \times 10^{B \times FWHM_{\rm corr}}\,{\rm M}_{\odot}$, where $B = 0.233 \pm 0.0041$ and $A = 1.81^{+1.05}_{-0.68} \times 10^{-3}$. The FWHM$_{\rm corr}$ of SN 2018zd at $t\approx399$\,d is measured to be  $54.2\pm1.0$\,\AA, which yields $M(^{56}{\rm Ni}) = 0.033\pm0.004\,{\rm M}_{\odot}$. This estimate is consistent with the value derived  from the light-curve tail (see Section \ref{subsect:Bolo}).

\section{Discussion}
\label{sect:Dis}

\begin{figure}
\centering
\includegraphics[width=8cm,angle=0]{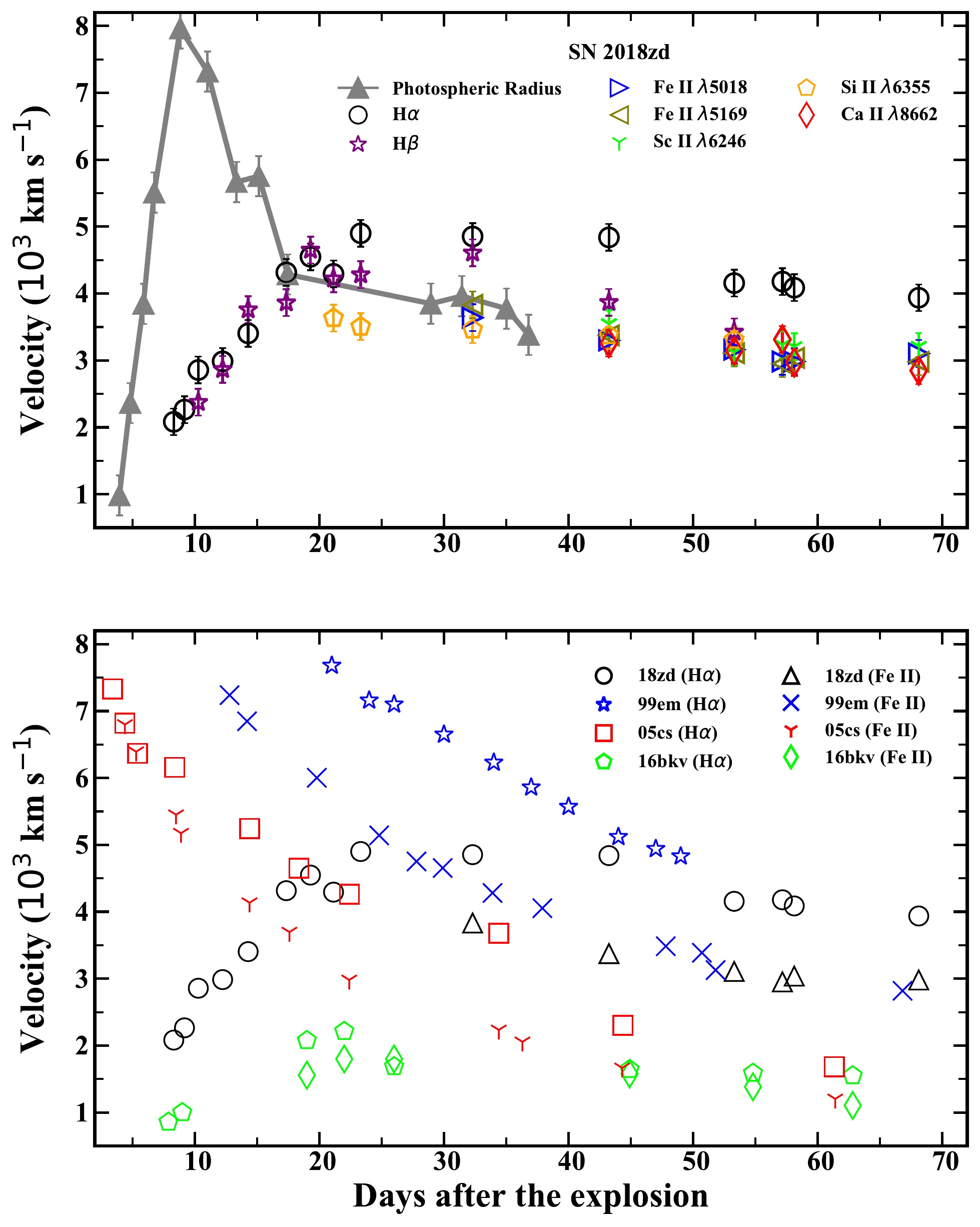}
 \caption{{\it Top:} Ejecta velocity of SN 2018zd derived from the absorption minima of H$\alpha$, H$\beta$, \ion{Fe}{ii} $\lambda$5018, \ion{Fe}{ii} $\lambda$5169, \ion{Sc}{ii} $\lambda$6246, \ion{Si}{ii} $\lambda$6355, and \ion{Ca}{ii} $\lambda$8662. The velocity derived from the photospheric radius presented in Fig. \ref{<SED>} is also plotted. {\it  Bottom:} Ejecta velocity of SNe 2018zd, 1999em, 2005cs, and 2016bkv derived from the absorption minima of H$\alpha$ and \ion{Fe}{ii} $\lambda$5169. The uncertainty is about 200--300\,km\,s$^{-1}$.}
\label{<vel>}
\end{figure}

\subsection{Ejecta velocities}
\label{subsect:vel}

The ejecta velocities of SN 2018zd measured from the absorption minima (including H$\alpha$, H$\beta$, \ion{Fe}{ii} $\lambda$5018, \ion{Fe}{ii} $\lambda$5169, \ion{Sc}{ii} $\lambda$6246, \ion{Si}{ii} $\lambda$6355, and \ion{Ca}{ii} $\lambda$8662) and the expanded photospheric radius are presented in the top panel of Fig. \ref{<vel>}. For comparison, the ejecta velocities of SNe 1999em, 1998S,  2005cs, and 2016bkv derived from absorption minima of H$\alpha$ and \ion{Fe}{ii} $\lambda$5169 are plotted in the bottom panel.

The velocity derived from the expanding photospheric radius of SN 2018zd  ($ v = \Delta R / \Delta t$)  is $\sim 1000$\,km\,s$^{-1}$ at $t\approx4$\,d, which accelerates at an average of 1200\,km\,s$^{-1}$\,d$^{-1}$ in the next five days. It reaches the maximum velocity ($\sim 8000$\,km\,s$^{-1}$) at $t\approx8.8$\,d when the bolometric luminosity rose to the peak. This velocity decreases quickly during the next week and reaches a velocity plateau at  $\sim 3600$\,km\,s$^{-1}$. The acceleration inferred from the photospheric radius might relate to the surrounding density structure of the SN. A CSM cloud with a density profile increasing outward may absorb progressively more energy from the shock, resulting in a fast-rising luminosity curve that may mimic an accelerating expansion of the photosphere.

The remarkable acceleration is seen in the velocity derived from the absorption minima of H$\alpha$ and H$\beta$ of SN 2018zd at $t \lesssim 20$\,d. At the same phase, the H$\alpha$ velocity of SN 2016bkv shows a similar increase, as noted by \cite{Nakaoka18}.  This acceleration might imply distinct line-forming regions. At early phases, the weak H absorption forms in the wind/CSM that is above the optically-thick photosphere. The shocked ejecta ionise and accelerate this material to yield the observed spectral lines and acceleration.  The H absorption lines in the well-developed P-Cygni profile at $t>20$\,d form in the ejecta that decelerate owing to obstruction by the outer material. 

At $t> 20$ d, some well-developed absorptions can be used to estimate the velocity of SN 2018zd besides of the H lines. The velocity derived from those absorptions at $t\approx30$ d is close to that from the photospheric radius but is  $\sim$ 1000 km s$^{-1}$ slower than that from H lines. As usually seen in the normal SNe IIP, the velocity derived from H lines is higher than those from the other ions \citep{Leonard02}. Considering these differences, we use  \ion{Fe}{ii}$\lambda$5169 and \ion{Sc}{ii}$\lambda$ 6246  as the tracker of photospheric velocity instead of H lines, as suggested by \cite{Hamuy03} and \cite{Maguire10}.  The velocity of SN 2018zd at $t = 50$ d interpolated by the velocity of \ion{Fe}{ii}$\lambda$5169 and \ion{Sc}{ii}$\lambda$ 6246 is about 3300 km s$^{-1}$, which is at the slow side of normal SNe IIP (as seen in Section \ref{sect:Diver}) and is faster than the low-velocity SNe IIP (e..g., SN 2005cs). 

\begin{figure}
\centering
\includegraphics[width=8.5cm,angle=0]{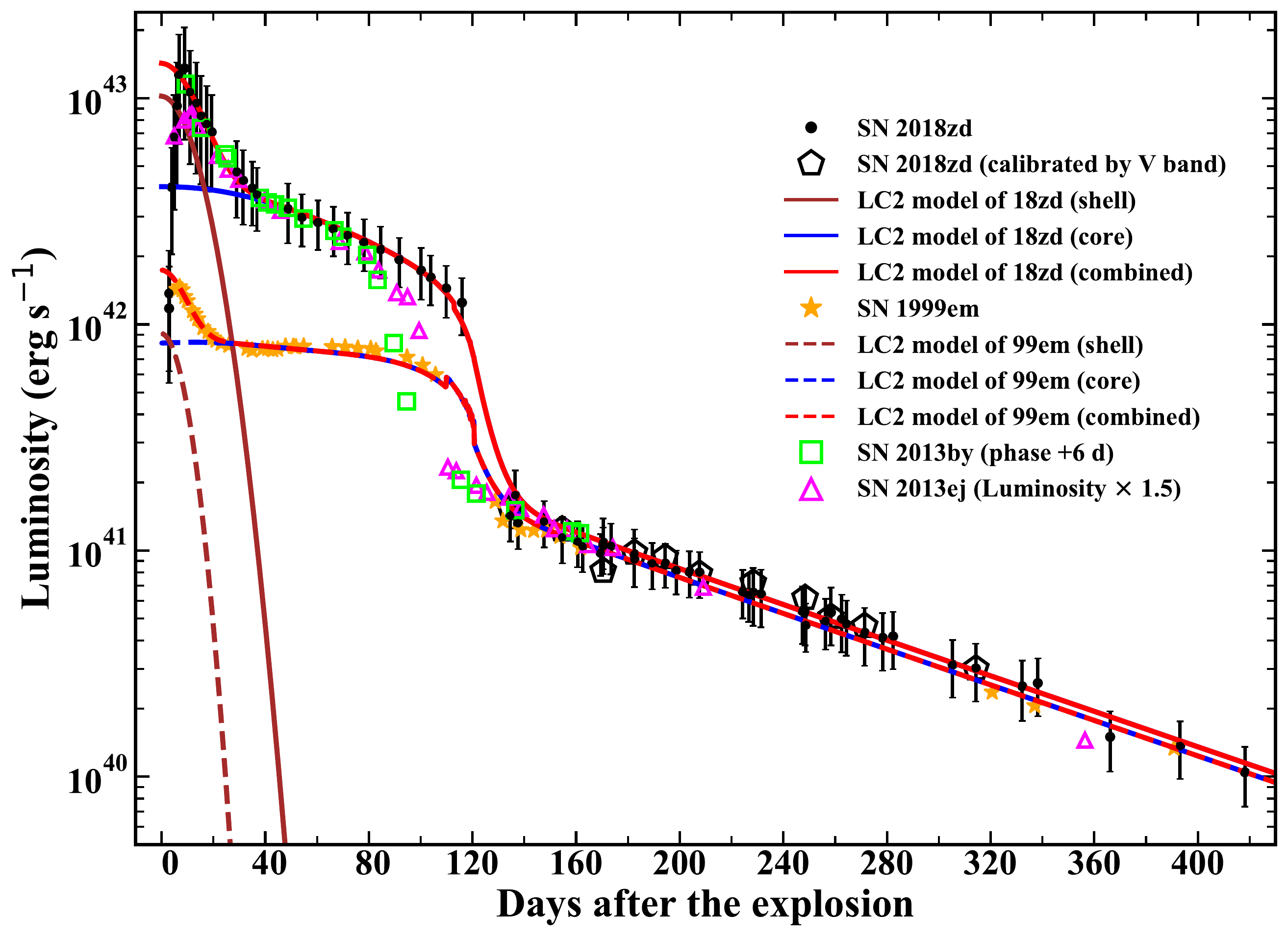}
\caption{Bolometric light curve of SN 2018zd compared with that of SN 1999em, SN 2013by, SN 2013ej and  the results of the LC2 model.  The error bars include uncertainties in the photometry, extinction, and distance, the last two of which greatly dominate. The LC2 model contains the contribution of the core and the shell. The phase of SN 2013by has been shifted by +6\,d corresponding to the archival explosion date (MJD = 56403.5; \citealp{Valenti15}). The luminosity of SN 2013ej has been multiplied by 1.5.}
\label{<bolo>}
\end{figure}

\begin{table*}
\small
\caption{Parameters of bolometric light curves$^a$}
\scalebox{0.93}{
\begin{tabular}{lccccccc}
\hline\hline
Supernova &Distance (Mpc)&E(B-V)$_{\rm total}$ & $L_{\rm peak} (10^{43} \rm erg s^{-1})$   &  $M(^{56}Ni) (\rm M_{\odot})^b$ &Explosion date (MJD) &Ref.$^c$ \\
\hline
SN 2018zd&18.4 (4.5)&0.17 (0.05)&1.36 (0.63)&0.035 (0.012)&58178.4 (0.5)&1\\
SN 1999em&8.2 (0.6)&0.10 (0.05)&0.15 (0.03) &0.033 (0.006)&51475.1 (1.4)&2\\
SN 2013by&14.8 (1.0)&0.20 (0.03)&1.15 (0.24)&0.036 (0.006)&56397.5 (2.0)&3,1\\
SN 2013ej&9.6 (0.7)&0.10 (0.03)&0.57 (0.08)&0.024 (0.005)&56496.5 (1.5)&4\\

\hline
\hline
\label{Tab:boloLC}
\end{tabular}
}\\
$^a${The parameters in Fig. \ref{<bolo>} . The number given in the brackets are the 1$\sigma$ uncertainties.  } $^b${.Mass of $^{56} \rm Ni$ derived by the bolometric luminosity during 130 d $\lesssim t \lesssim$ 180 d via Eq. \ref{eq:Ni}. } $^c${Reference of the distance, reddening and explosion date adopted in the calculation: 1. This work,  2. \cite{Leonard02}, 3. \cite{Valenti15}, 4. \cite{Huang15}.  }

\end{table*}

\subsection{Bolometric flux and explosion parameters}
\label{subsect:Bolo}
Fig. \ref{<bolo>} displays the bolometric light curve of SN 2018zd based on the SED presented in Section \ref{subsect:SED}. The flux at wavelengths redder than the photometry is derived by extrapolating a blackbody fit to the SED. The flux at wavelengths shorter than 1216\,\AA\, is assumed to be zero because of absorption in the H-rich atmosphere (Lyman series). At $t < 115$\,d, when  UV photometry is available, we extrapolate the $uvw2$ band (or the $uvm2$ band if $uvw2 - u \gtrsim 1.0$\,mag to avoid the red leak of the $uvw2$ band; \citealp{Brown16}) flux using a straight line to zero flux at 1216\,\AA\, (as seen in Fig. \ref{<SED>}).  In the tail phase, we omit the flux at wavelengths shorter than 3600\,\AA\, and extrapolate the flux beyond the wavelength range covered by the photometry via the blackbody directly.  The latter treatment yields a luminosity match to that derived using the $V$-band photometryvia the equation \citep{Bersten09}\footnote{The ``$-11.64$'' in Equation 4 of \cite{Bersten09} is altered to be ``+11.64'' to correct the typo.}  during 135 d $\lesssim t \lesssim$ 338 d:  
\begin{equation}
{\rm log}\,L = -0.4[BC + V - A_V + 11.64] + {\rm log}\,(4 \pi D^2),
\label{eq:V_cor09}
\end{equation}
\noindent
where $L$ is given in units of erg\,s$^{-1}$, the colour term $BC = -0.7$\,mag, and $D$ is the SN distance in units of cm.

The bolometric light curves of SN 1999em, SN 2013by, and SN 2013ej produced by the same method, and the bolometric light curve generated by the two-component model (LC2) of  \cite{Nagy16}, are also shown for comparison.  LC2  is a semianalytical light-curve model, which computes the bolometric light curve of a homologous expanding supernova using the radiative diffusion approximation as introduced by \cite{Arnett89}.

SN 2018zd reached its bolometric peak at $t \approx 8.8$\,d, $L_{\rm max} = (1.36\pm0.63) \times 10^{43}$\,erg\,s$^{-1}$, higher than that of SN 1999em by a factor of eight, as listed in Table \ref{Tab:boloLC}. The large peak luminosity of SN 2018zd is partly due to the contribution of interaction between the shocked ejecta and stellar wind/CSM. This SN reached UV and optical maxima almost simultaneously, which also helped explain the higher peak bolometric luminosity. At the tail phase ($t \gtrsim 130$\,d), however, these two SNe are found to have  similar luminosity ($L \approx 1.2 \times 10^{41}$\,erg\,s$^{-1}$ at $t \approx 150$\,d) and decline rate ($\sim 0.98$\,mag\,100\,d$^{-1}$, as expected  from Co$\rightarrow$Fe decay).

The expansion velocity of SN 2018zd matches that of SN 1999em within  $\sim 200$\,km\,s$^{-1}$ at $t \approx$ 50\,d, as presented in Fig. \ref{<vel>}. They are expected to have a similar middle plateau luminosity at a statistical uncertainty of 0.5\,mag if they follow the velocity-luminosity relation \citep{Hamuy03}, as seen in panel (a) of Fig. \ref{<diver>}.  Moreover, SN 2018zd exhibits a tail similar to that of SN 1999em after the contribution of CSM interaction disappeared. Thus, it might be reasonable to assume that the energy produced in the explosion of SN 2018zd is close to that of SN 1999em at the uncertain level of the velocity-luminosity relation if the former has no extra energy input from the interaction. Assuming all of the excess emission inferred in SN 2018zd relative to that of SN 1999em is converted from the kinetic energy of the ejecta ($M_{\rm wind}=\int_{3}^{120} [{L_{\rm 18zd}(t) - L_{\rm 99em}(t)}]/{v_{\rm ej}^2(t)}\,dt$, where the photospheric ejecta velocity $v_{\rm ej}(t)$ is derived from the absorption minimum of \ion{Fe}{ii} lines, as presented in Fig. \ref{<vel>}; $v_{\rm ej}(t)$ at $t > 70$\,d is set to 3300\,km\,s$^{-1}$, the integration limit is from the first date when SN 2018zd having multi-band photometry to the end of plateau phase), we find that $0.18^{+0.05}_{-0.10}\,{\rm M}_{\odot}$ of wind material is required to produce the extra flux seen in SN 2018zd. The uncertainty in the wind-mass estimate includes the uncertainties in bolometric flux and the velocity-luminosity relation. This stellar wind is more massive than that of SN 2013fs ($\lesssim 0.01\, {\rm M}_{\odot}$; estimated via the line luminosity of the narrow H$\alpha$ emission in \citealp{Yaron17}, and spectral modeling in \citealp{Dessart17}) but less massive than that of SN 1998S ($\sim0.4\,{\rm M}_{\odot}$; derived from  spectral modeling in \citealp{Dessart16}).

SN 2013by shows a bolometric light curve similar to that of SN 2018zd at $t \lesssim 80$\,d if its epoch is shifted by +6\,d. Given that the closest available prediscovery image of SN 2013by is about 22\,d earlier than the discovery image \citep{Parker13} it is possible that SN 2013by exploded a few days earlier  than the explosion date reported by \cite{Valenti15}.  Thus, we adopt MJD = 56397.5 as the explosion date of SN 2013by for the following calculation. With this modified explosion date, the colour curve and light curve of SN 2013by matches that of SN 2018zd better. At $t > 80$\,d, SN 2013by ends the plateau phase and fades quickly. However, its tail luminosity meets that of SN 2018zd again.

The bolometric light curve of SN 2013ej roughly matches that of SN 2018zd and SN 2013by except the middle part (80\,d $\lesssim t \lesssim$ 120\,d) if the luminosity of SN 2013ej is multiplied by 1.5. These three SNe show a faster-declining plateau and a more significant drop from the plateau to the radioactive tail than does SN 1999em.  

We estimate the mass of $^{56}$Ni ejected by the SN based on the  tail  bolometric luminosity via \citep{Hamuy03}:
\begin{equation}
M_{\rm Ni} = (7.866\times10^{-44}) L_t \times {\rm exp}\left[ \frac{t/(1+z)-6.1}{111.26} \right]\,M_{\sun},
 \label{eq:Ni}
\end{equation}
where $t$ is the phase after explosion, $L_t$ is the tail-phase luminosity in units of erg\,s$^{-1}$, and $z$ is the redshift of the SN.  SN 2018zd, SN 1999em, and SN 2013by produced similar amounts of $^{56}$Ni, as listed in Table \ref{Tab:boloLC}, given the similar radioactive tails.

\begin{table}
\centering
\caption{LC2 model parameters of SN\,2018zd and SN\,1999em$^a$}
\begin{tabular}{lcccc}
\hline\hline
Parameter & SN 2018zd & & SN 1999em &  \\
 & ``core'' & `shell''  & ``core'' & ``shell'' \\
\hline
Initial & model& parameters\\
\hline
$R_0$ ($10^{13}$\,cm)	&	4.8	&	7.0	&	3.5	&	5.0	\\
$M_{\rm{ej}}$  (M$_{\odot}$)	&	9.80	&	0.55	&	13.50	&	0.20	\\
$M_{\rm Ni}$\, (M$_{\odot}$)	&	0.033	&	0	&	0.030	&	0	\\
$T_{\rm rec}$ (K)	&	6000	&	0	&	7000	&	0	\\
$E_{\rm kin}$ ($10^{51}$\,erg)	&	2.20	&	1.30	&	1.48	&	0.40	\\
$E_{\rm th}$ ($10^{51}$\,erg)	&	1.90	&	0.40	&	0.40	&	0.02	\\
$\alpha$	&	1.6	&	0.0	&	0.0	&	0.0	\\
$\kappa$ (cm$^2$\,g$^{-1}$)	&	0.28	&	0.40	&	0.26	&	0.40	\\
\hline
Calculated & physical & properties\\
\hline
$t_0$(d)	&	95.6	&	18.7	&	102.9	&	11.7	\\
$v$ (km\,s$^{-1}$)	&	6580	&	23200	&	4290	&	21300 \\
\hline
\end{tabular}
$^a${$R_0$ is the initial radius of the ejecta, $M_{\rm{ej}}$ is the ejected mass, $M_{\rm Ni}$ is the initial nickel mass, $T_{\rm rec}$ is the recombination temperature, $E_{\rm kin}$ is initial kinetic energy,  $E_{\rm th}$ is the initial thermal energy, $\alpha$ is the density profile exponent, $\kappa$ is the opacity, $t_0$ is the light-curve timescale, and $v$ is the maximum expansion velocity.}

\label{Tab:LC2}
\end{table}

It is not surprising to see that the mass of $^{56}$Ni derived by the LC2 model, $0.033\,{\rm M}_{\odot}$), is consistent with the above results, because they rely on a shared physical basis \citep{Arnett82,Arnett89}. Thus, the mean value of $M(^{56}$Ni) derived via the tail-phase flux matches the result from the nebular spectra. Since the estimate from the nebular spectra is relies only on the FWHM of H$\alpha$ emission, which is independent of distance and extinction, the agreement between these two methods suggests that the distance and extinction of SN 2018zd adopted in this paper are reasonable. 

The LC2 model reproduces the light curve of SN 1999em and also the light curve of SN 2018zd at $t \gtrsim 10$\,d with the parameters listed in Table \ref{Tab:LC2}. Note that all models shown are not from formal fitting to the data. Instead, they can be considered only as representative examples that look similar to the observations. Thus, formal uncertainties cannot be assigned to the parameters shown in Table \ref{Tab:LC2}, but they are at least $\sim 10$ per cent according to \cite{Nagy16}.

The velocity derived in the LC2 model is about 2000\,km\,s$^{-1}$ higher than that of SN 2018zd obtained from the spectral features, as presented in Section \ref{subsect:vel}. This suggests that extra energy is required to produce such a luminous SN in the scheme of regular SNe~IIP. A significant amount of the SN kinetic energy may be converted to thermal energy via the CSM interaction, conforming to the high temperature seen in Fig. \ref{<SED>}.

\subsection{Comparison with other SNe II}
\label{sect:Diver}

\begin{figure*}
\centering
\includegraphics[width=18cm,angle=0]{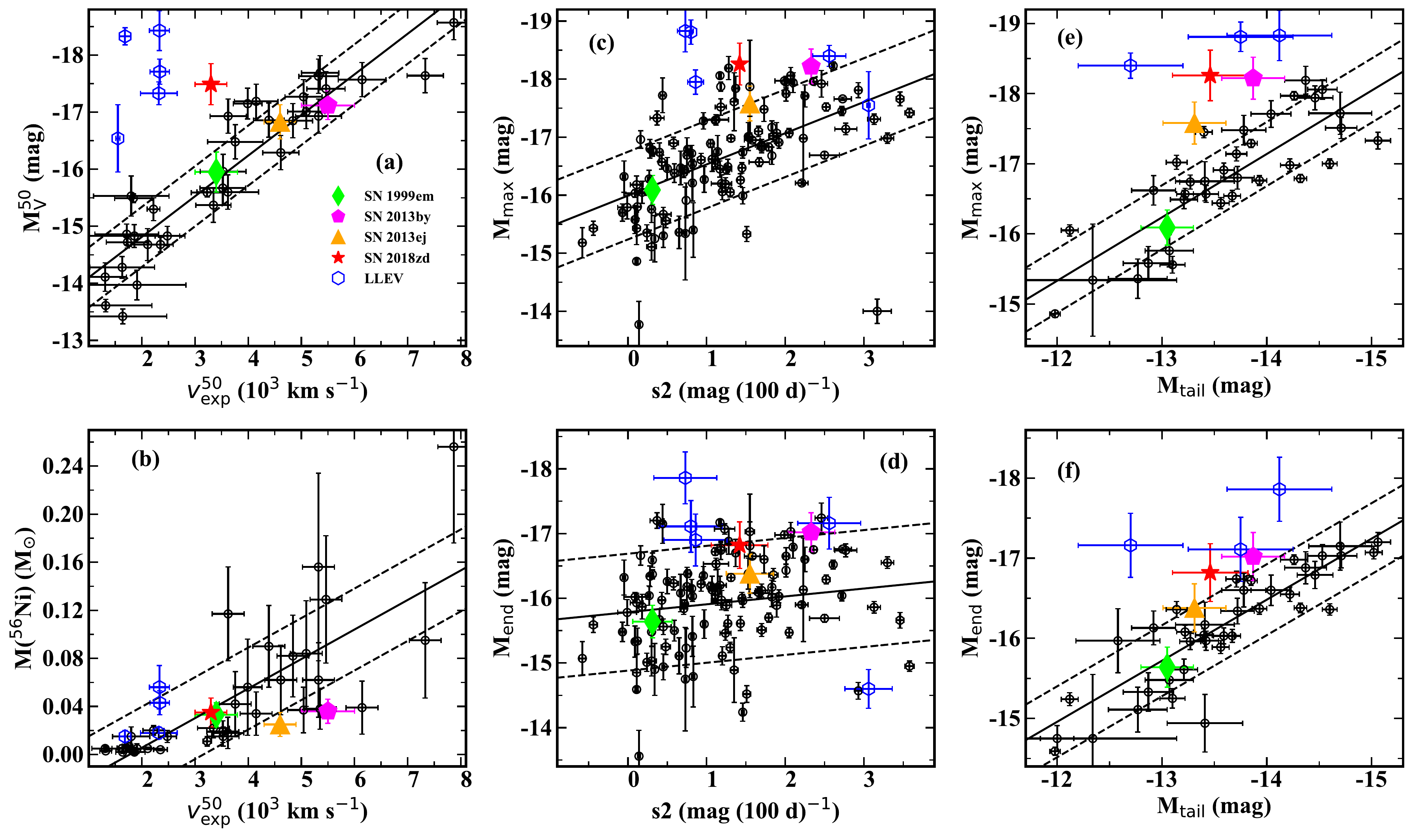}
 \caption{The position of SN 2018zd in the SN~II family considering various photometric and spectroscopic indicators, including the velocity of \ion{Sc}{ii} $\lambda$6246 or \ion{Fe}{ii} $\lambda$5169 measured at $t\approx 50$\,d after explosion (i.e., $v^{50}_{\rm exp}$), the $V$-band absolute magnitude measured at $t\approx 50$\,d after explosion (i.e., $M_V^{50}$), the mass of $^{56}$Ni (i.e., $M(^{56}$Ni)), and four shape parameters of the $V$-band light curve (as defined by \citealp{Anderson14}) --- the absolute brightness at three selected phases (i.e.,  $M_{\rm max}$, $M_{\rm end}$, and $M_{\rm tail}$) and the decline rate of the plateau (i.e., $s2$). The solid lines are the linear fits  to black open circles (collected from \citealp{Hamuy03,Anderson14,Spiro14,Zhang14}), and the dashed lines are the 1$\sigma$ uncertainties of the linear fits.  SN 1999em, SN 2013ej, SN 2013by, and five luminous SNe with low expansion velocities \citep[LLEV~SNe;][]{Rodriguez20} are included for comparison.}
\label{<diver>}
\end{figure*}

Fig. \ref{<diver>} displays the position of SN 2018zd in the spectroscopic and photometric parameter space of SNe~II. Comparing with the linear fitting in each panel, we note the following:\\

(1) The peak ($M_{\rm max}$) and middle plateau ($M_V^{50}$) of SN 2018zd are (respectively) 1.50 $\pm 0.75$\,mag and $1.75 \pm 0.53$\,mag  brighter than the sample SNe having a similar decline rate $s2$ \citep{Anderson14} and ejecta expansion velocity ($v^{50}_{\rm exp}$) in panels (a) and (c).\\

(2)  SN 2018zd experienced a significant luminosity drop from the peak to the tail, a $\sim 2 \sigma$ departure from the linear fitting of $M_{\rm max}$ vs. $M_{\rm tail}$ in panel (e).\\

(3) The tail brightness of SN 2018zd follows that of the SN~II family.  For example, the mass of synthesised $^{56}$Ni in SN 2018zd matches the 
$v^{50}_{\rm exp}$ vs. $M(^{56}$Ni) relation in panel (b), and the brightness at the beginning of the tail phase is located in the $1\sigma$ region of $s2$ vs. $M_{\rm end}$ space in panel (d).\\

Two other SNe~II, SN 2013ej and SN 2013by, show somewhat similar behaviour not only in bolometric light curves but also their positions in the SN~II family. They have an intermediate flux drop during the transition from the plateau to the radioactive tail, smaller than that of SN 2018zd but more significant than the sample cluster. Combined with their fast light-curve decline rate during the photospheric phase, these two SNe might have extra energy injected by CSM interaction. \cite{Valenti15} also suggest that the interaction scenario explains the fast-declining light curves of SN 2013by based on the flash spectral features in the spectrum at $t\approx$2\,d (as seen in Fig. \ref{<Sp_comp>}) and the detection of the X-ray emission. Moreover, \cite{Morozova17} use a dense wind-like CSM interaction model to produce the luminous and fast-declining light curve of SN 2013ej and SN 2013by. The brightness of SN 2013ej and SN 2013by in the middle of the plateau phase follows the velocity-luminosity relation of SNe~IIP, as seen in the panel (a) of Fig. \ref{<diver>}. It seems that the extra energy in these two SNe disappeared at $t\approx50$\,d, which suggests an interaction lasting for a period shorter  than  that of SN 2018zd. 

SN 2018zd is an outlier in velocity-brightness space, as seen in panel (a). However, a group of luminous SNe with low expansion velocities (LLEV~SNe; \citealp{Rodriguez20}) is located in a more distant region relative to the bulk of sample than SN  2018zd. The middle plateau brightness of the LLEV~SNe is $\sim 2$--3\,mag brighter than that of SNe~II having the same ejecta expansion velocity.  A larger drop from the peak to the tail in the light curve is also found in the LLEV~SNe.  \cite{Rodriguez20} listed the observational characteristics of the LLEV~SNe that are also found in SN 2018zd, such as the ejecta-CSM interaction signs at early phases, blue $B-V$ colours, weakness of metal lines, and luminous peak and plateau compared with the low expansion velocities. All of these properties were reproduced, assuming ejecta-CSM interaction that lasts between 4 and 11 weeks post-explosion, in the model work of \cite{Rodriguez20}. It confirms the interaction scenario of SN 2018zd suggested by our analysis here.

\subsection{Electron-capture supernova}
\label{sect:ecSNe}

The typical temperature of the shocked CSM is $\sim 10^9$\,K and that of the shocked ejecta is $\geq 10^7$\,K, and both radiate in the X-ray band \citep{Branch17}. These X-ray photons ionise the atoms in the surrounding wind/CSM, which then radiate narrow emission lines in the UV-optical band (i.e., the flashed spectral features) owing to recombination. Thus, we expect to detect the X-rays from SNe~II if this radiation is not strongly absorbed by the unshocked material. However,  \cite{Chandra18} reported the nondetection of SN 2018zd with the {\it Swift} X-ray telescope (XRT) between March 04 and March 11, 2018. They found an upper flux limit of $5.22 \times 10^{-14}$\,erg\,cm$^2$\,s$^{-1}$ (0.3--10\,keV).  Adopting $D = 18.4$\,Mpc, the upper limit for the X-ray luminosity of SN 2018zd is $\sim 2.11 \times 10^{39}$\,erg\,s$^{-1}$, slightly less than the detected luminosity of SN 2013by ($\sim 2.88 \times 10^{39}$\,erg\,s$^{-1}$; \citealp{Valenti15}).  The weaker X-ray radiation of SN 2018zd than that of SN 2013by might be related to the stronger flash emission features of the former.  It is possible that the majority of the X-ray photons from the shocked material are absorbed by the wind/CSM, producing stronger emission lines in the optical spectra.

\cite{Moriya14} suggested that the optically bright but X-ray-faint SNe IIn can arise from electron-capture supernovae (ECSNe) that is exploded from the super-asymptotic giant branch (super-AGB) star \citep{Miyaji80,Nomoto82,Miyaji87}. Super-AGB stars have a dense and massive circumstellar environment due to its high mass-loss rate wind, which could create narrow emission lines during the SN explosion, as seen in SN 2018zd. ECSNe is possible to be as luminous as SNe IIn at the peak phase due to the larger radius of the progenitor star, but the tail is faint because of the intrinsic low explosion energy and a small amount of $^{56}$Ni \citep{Tominaga13,Moriya14}. The synthesised mass of $^{56}$Ni in SN 2018zd is relatively small compared to its luminous peak. Furthermore, this SN produced even less $^{56}$Ni if a shorter distance is adopted in the calculation. For example, it yields only $0.013 \pm 0.004$\,M$_{\odot}$ if $D = 11.3 \pm 2.0$\,Mpc \citep{Bottinelli84}, which is close to the upper limit mass of $^{56}$Ni produced in ECSNe ($ < 0.015 \rm M_{\odot}$; \citealp{Kitaura06}). 

Thus, given the luminous UV-optical peak,  the relatively weak X-ray radiation, the massive wind/CSM environment, and a relatively small amount of synthesised $^{56}$Ni,  the electron-capture trigged supernova explosion is also a possible  scenario of SN 2018zd.

\section{Conclusions}
\label{sect:sum}

We present an extensive set of observations of SN 2018zd from the early ``flash-ionisation'' spectroscopic phase (e.g., $t \lesssim 1$\,d) to the later nebular phase (e.g., $t\approx456$\,d). The flux excess observed in the clear-band light curve $\sim 3.6$\,hr after the explosion might be a signal of shock breakout. The overluminous peak [e.g., $L_{\rm max}=(1.36\pm0.62)\times10^{43}$\,erg\,s$^{-1}$] of this SN accompanied by flash spectra at early times is likely caused by extra energy injection via interaction between the shocked ejecta and the massive wind ($0.18^{+0.05}_{-0.10}\,{\rm M}_{\odot}$). This effect does not disappear immediately when the flash signatures in the spectra fade away. Energy is stored in the H-rich envelope and heats it, making SN 2018zd bluer than regular SNe~IIP, as seen in its colour curve and spectrum during the photospheric phase. Meanwhile, this energy makes a brighter mid-plateau for SN 2018zd than regular SNe~II having the same expansion velocity. SN 2018zd is thus a gap-filler event between the central cluster and separated LLEV~SNe in velocity-brightness space, perhaps implying a continuous distribution of the interaction contribution. 
Given the similar fast-declining light curve of SNe~II reported by \cite{Huang15} for SN 2013ej and \cite{Valenti15} for SN 2013by, there is a group of transitional events located at the boundary between SNe~IIP and SNe~IIL in the sense of decline rate.  
Finally, the nebular spectra and the light-curve tail of SN 2018zd resemble those of typical SNe~IIP when the injected energy has been released. Thus, the observations of SN 2018zd presented in this paper are key to understanding the nature of the diverse origins of SNe~II.

\section*{Acknowledgements}
An anonymous referee made many useful suggestions that improved this paper. We acknowledge the support of the staff of the LJT, TNT, XLT, KOT, WT, and Keck. Funding for the LJT has been provided by the CAS and the People's Government of Yunnan Province. The LJT is jointly operated and administrated by YNAO and Center for Astronomical Mega-Science, CAS. Some of the data presented herein were obtained at the W. M. Keck Observatory, which is operated as a scientific partnership among the California Institute of Technology, the University of California, and NASA; the Observatory was made possible by the generous financial support of the W. M. Keck Foundation. We are grateful for discussions with Luc Dessart, Osmar Rodr\'iguez, Mario Hamuy, Peter Brown, Paolo Mazzali, and Pina Elena about this work.

J.Z. is supported by the National Natural Science Foundation of China (NSFC, grants 11773067, 11403096), by the Western Light Youth Project, and by the Youth Innovation Promotion Association of the CAS (grant 2018081). This work is sponsored (in part) by the CAS South America Center for Astronomy (CASSACA) in Santiago, Chile.
X.W. is supported by NSFC (grants 11633002 and 11761141001) and the Major State Basic Research Development Program (grant 2016YFA0400803).
J.V. and his group at Konkoly Observatory are supported by the project ``Transient Astrophysical Objects'' GINOP 2.3.2-15-2016-00033 of the National Research, Development and Innovation Office (NKFIH), Hungary, funded by the European Union. R.K.-T. is also supported by the \'UNKP-19-02 New National Excellence Program of the Ministry for Innovation and Technology. {\L}.W. acknowledges support from the the Polish NCN DAINA grant 2017/27/L/ST9/03221 and EC's H2020 OPTICON grant 730890.
A.V.F.'s supernova group is grateful for financial assistance from the Christopher R. Redlich Fund, the TABASGO Foundation, and the Miller Institute for Basic Research in Science (U.C. Berkeley).
P.M. acknowledges support provided by the Polish National Science Center (NCN) through grant 2016/21/B/ST9/01126. We acknowledge the Gaia Photometric Alerts group and the use of the Cambridge Photometric Calibration Server (http://gsaweb.ast.cam.ac.uk/followup), developed by Sergey Koposov and maintained by {\L}ukasz Wyrzykowski, Arancha Delgado, and Pawel Zielinski, funded by the European Union's Horizon 2020 research and innovation programmer under grant agreement 730890 (OPTICON). K.V. and L.K. are grateful for financial support from the National Research, Development and Innovation Office (NKFIH), Hungary, under grants NKFI-K-131508 and NKFI-KH-130526. \'A.S., A.B., C.K., and K.V. are supported by the Lend\"ulet program grant LP2018-7/2019 of the Hungarian Academy of Sciences. 








\appendix

\section{Photometric and spectroscopic data}

\begin{table*}
\scriptsize
\caption{Local photometric standard stars in the field of SN\,2018zd$^a$}
\begin{tabular}{ccccccccc}
\hline\hline
Star &  $U$ (mag) & $B$ (mag) & $V$ (mag) & $R$ (mag) & $I$ (mag) & $g$ (mag)& $r$ (mag) & $i$ (mag) \\
\hline
1	&	14.88(0.01)	&	14.63(0.01)	&	13.79(0.01)	&	13.40(0.02)	&	13.02(0.02)	&	14.17(0.01)	&	13.56(0.01)	&	13.40(0.01)	\\
2	&	15.47(0.01)	&	14.92(0.01)	&	13.98(0.01)	&	13.53(0.03)	&	13.12(0.01)	&	14.42(0.01)	&	13.68(0.01)	&	13.47(0.01)	\\
3	&	16.35(0.02)	&	16.47(0.01)	&	15.93(0.02)	&	15.58(0.02)	&	15.26(0.02)	&	16.15(0.01)	&	15.78(0.01)	&	15.67(0.01)	\\
4	&	16.49(0.01)	&	16.32(0.01)	&	15.54(0.03)	&	15.10(0.01)	&	14.69(0.01)	&	15.90(0.01)	&	15.31(0.01)	&	15.11(0.01)	\\
5	&	17.54(0.04)	&	17.26(0.03)	&	16.41(0.03)	&	15.96(0.04)	&	15.53(0.02)	&	16.79(0.01)	&	16.18(0.01)	&	15.96(0.01)	\\
6	&	15.48(0.02)	&	15.48(0.02)	&	14.80(0.02)	&	14.41(0.01)	&	14.04(0.01)	&	15.09(0.01)	&	14.61(0.01)	&	14.46(0.01)	\\
7	&	17.08(0.01)	&	16.86(0.02)	&	16.06(0.02)	&	15.61(0.01)	&	15.19(0.02)	&	16.42(0.01)	&	15.82(0.01)	&	15.61(0.01)	\\
8	&	17.48(0.02)	&	16.86(0.02)	&	15.92(0.01)	&	15.45(0.02)	&	15.03(0.02)	&	16.35(0.02)	&	15.61(0.01)	&	15.40(0.02)	\\
9	&	17.94(0.01)	&	17.89(0.01)	&	17.08(0.01)	&	16.59(0.02)	&	16.13(0.02)	&	17.45(0.01)	&	16.83(0.01)	&	16.59(0.01)	\\
10	&	14.56(0.01)	&	14.44(0.02)	&	13.72(0.01)	&	13.40(0.01)	&	13.08(0.01)	&	14.02(0.01)	&	13.56(0.01)	&	13.45(0.01)	\\
11	&	17.26(0.03)	&	17.07(0.01)	&	16.27(0.01)	&	15.85(0.02)	&	15.46(0.03)	&	16.63(0.01)	&	16.04(0.01)	&	15.85(0.01)	\\
12	&	16.85(0.01)	&	16.52(0.01)	&	15.67(0.01)	&	15.24(0.01)	&	14.83(0.03)	&	16.06(0.01)	&	15.42(0.01)	&	15.23(0.01)	\\
\hline
\hline
\end{tabular}

$^a${See Fig. \ref{<img>} for the finder chart of these reference stars. $UBVRI$ bands in Vega magnitude system, $gri$ bands in AB magnitude system. Uncertainties (in parentheses) are $1\sigma$.}

\label{Tab:Photo_stand}
\end{table*}

\begin{table*}
\scriptsize
\caption{Ground-based photometry of SN 2018zd$^a$}
\begin{tabular}{lccccccccccc}
\hline
\hline
Date (UT) & MJD & Epoch (d)$^b$ & $U$ (mag) & $B$ (mag) & $V$ (mag) & $R$ (mag)& $I$ (mag)& $g$ (mag) & $r$ (mag)& $i$ (mag) & Facility\\
\hline
Mar. 04 2018	&	58181.91	&	3.52	&	...	&	15.38(0.01)	&	15.26(0.01)	&	15.05(0.01)	&	14.91(0.01)	&	...	&	...	&	...	&	KOT	\\
Mar. 05 2018	&	58182.76	&	4.37	&	...	&	14.93(0.01)	&	14.73(0.01)	&	...	&	...	&	14.79(0.01)	&	14.76(0.01)	&	14.87(0.01)	&	LJT	\\
Mar. 07 2018	&	58184.70	&	6.31	&	13.30(0.02)	&	14.07(0.02)	&	13.91(0.01)	&	...	&	...	&	13.94(0.01)	&	13.96(0.01)	&	14.10(0.01)	&	LJT	\\
Mar. 07 2018	&	58184.73	&	6.34	&	...	&	13.85(0.08)	&	13.96(0.05)	&	13.77(0.08)	&	13.73(0.05)	&	...	&	...	&	...	&	KOT	\\
Mar. 08 2018	&	58185.48	&	7.09	&	...	&	13.93(0.01)	&	13.67(0.01)	&	13.59(0.01)	&	13.57(0.01)	&	...	&	...	&	...	&	TNT	\\
Mar. 08 2018	&	58185.70	&	7.31	&	13.03(0.03)	&	13.82(0.02)	&	13.65(0.01)	&	...	&	...	&	13.68(0.01)	&	13.70(0.01)	&	13.85(0.02)	&	LJT	\\
Mar. 08 2018	&	58185.75	&	7.36	&	...	&	13.70(0.05)	&	13.72(0.04)	&	13.52(0.10)	&	13.45(0.09)	&	...	&	...	&	...	&	KOT	\\
Mar. 09 2018	&	58186.47	&	8.08	&	...	&	13.81(0.05)	&	13.51(0.10)	&	13.45(0.08)	&	13.43(0.09)	&	...	&	...	&	...	&	TNT	\\
Mar. 09 2018	&	58186.70	&	8.31	&	12.98(0.03)	&	13.79(0.02)	&	13.55(0.01)	&	...	&	...	&	13.63(0.01)	&	13.62(0.01)	&	13.78(0.01)	&	LJT	\\
Mar. 10 2018	&	58187.48	&	9.09	&	...	&	13.81(0.04)	&	13.49(0.04)	&	13.44(0.07)	&	13.40(0.04)	&	...	&	...	&	...	&	TNT	\\
Mar. 10 2018	&	58187.79	&	9.40	&	...	&	13.75(0.03)	&	13.63(0.02)	&	13.46(0.03)	&	13.36(0.04)	&	...	&	...	&	...	&	KOT	\\
Mar. 11 2018	&	58188.70	&	10.31	&	13.08(0.03)	&	13.86(0.01)	&	13.58(0.01)	&	...	&	...	&	13.69(0.01)	&	13.64(0.02)	&	13.74(0.01)	&	LJT	\\
Mar. 12 2018	&	58189.47	&	11.08	&	...	&	13.93(0.04)	&	13.55(0.04)	&	13.46(0.09)	&	13.37(0.04)	&	...	&	...	&	...	&	TNT	\\
Mar. 12 2018	&	58189.72	&	11.33	&	13.16(0.04)	&	13.94(0.03)	&	13.63(0.01)	&	...	&	...	&	13.75(0.01)	&	13.68(0.01)	&	13.79(0.01)	&	LJT	\\
Mar. 12 2018	&	58189.74	&	11.35	&	...	&	13.76(0.07)	&	13.65(0.04)	&	13.46(0.07)	&	13.36(0.07)	&	...	&	...	&	...	&	KOT	\\
Mar. 12 2018	&	58189.92	&	11.53	&	...	&	13.86(0.03)	&	13.71(0.02)	&	13.50(0.05)	&	13.36(0.07)	&	...	&	...	&	...	&	KOT	\\
Mar. 13 2018	&	58190.64	&	12.25	&	13.21(0.02)	&	13.95(0.02)	&	13.64(0.01)	&	...	&	...	&	13.78(0.02)	&	13.68(0.01)	&	13.78(0.01)	&	LJT	\\
Mar. 13 2018	&	58190.81	&	12.42	&	...	&	13.92(0.02)	&	13.76(0.01)	&	13.55(0.03)	&	13.39(0.04)	&	...	&	...	&	...	&	KOT	\\
Mar. 14 2018	&	58191.77	&	13.38	&	...	&	14.00(0.02)	&	13.78(0.01)	&	13.56(0.03)	&	13.41(0.04)	&	...	&	...	&	...	&	KOT	\\
Mar. 15 2018	&	58192.48	&	14.09	&	...	&	14.06(0.05)	&	13.65(0.04)	&	13.53(0.08)	&	13.46(0.05)	&	...	&	...	&	...	&	TNT	\\
Mar. 18 2018	&	58195.47	&	17.08	&	...	&	14.19(0.04)	&	13.75(0.04)	&	13.57(0.07)	&	13.47(0.08)	&	...	&	...	&	...	&	TNT	\\
Mar. 18 2018	&	58195.73	&	17.34	&	13.46(0.04)	&	14.18(0.01)	&	13.81(0.01)	&	...	&	...	&	13.95(0.01)	&	13.79(0.02)	&	13.85(0.01)	&	LJT	\\
Mar. 19 2018	&	58196.48	&	18.09	&	...	&	14.21(0.05)	&	13.75(0.10)	&	13.59(0.07)	&	13.47(0.09)	&	...	&	...	&	...	&	TNT	\\
Mar. 21 2018	&	58198.49	&	20.10	&	...	&	14.29(0.04)	&	13.83(0.05)	&	13.62(0.04)	&	13.48(0.06)	&	...	&	...	&	...	&	TNT	\\
Mar. 21 2018	&	58198.56	&	20.17	&	13.55(0.03)	&	14.22(0.01)	&	13.83(0.01)	&	...	&	...	&	13.99(0.01)	&	13.81(0.01)	&	13.84(0.01)	&	LJT	\\
Mar. 22 2018	&	58199.48	&	21.09	&	...	&	14.29(0.04)	&	13.83(0.03)	&	13.63(0.05)	&	13.48(0.04)	&	...	&	...	&	...	&	TNT	\\
Mar. 22 2018	&	58199.78	&	21.39	&	...	&	14.25(0.02)	&	13.97(0.01)	&	13.68(0.03)	&	13.47(0.03)	&	...	&	...	&	...	&	KOT	\\
Mar. 23 2018	&	58200.47	&	22.08	&	...	&	14.33(0.05)	&	13.85(0.06)	&	13.65(0.07)	&	13.50(0.05)	&	...	&	...	&	...	&	TNT	\\
Mar. 23 2018	&	58200.61	&	22.22	&	13.66(0.03)	&	14.30(0.01)	&	13.89(0.01)	&	...	&	...	&	14.06(0.01)	&	13.83(0.02)	&	13.86(0.01)	&	LJT	\\
Mar. 23 2018	&	58200.76	&	22.37	&	...	&	14.30(0.03)	&	13.99(0.02)	&	13.71(0.03)	&	13.51(0.03)	&	...	&	...	&	...	&	KOT	\\
Mar. 23 2018	&	58200.87	&	22.48	&	...	&	14.24(0.05)	&	13.93(0.04)	&	13.66(0.06)	&	13.46(0.06)	&	...	&	...	&	...	&	KOT	\\
Mar. 24 2018	&	58201.49	&	23.10	&	...	&	14.37(0.04)	&	13.87(0.05)	&	13.66(0.03)	&	13.49(0.04)	&	...	&	...	&	...	&	TNT	\\
Mar. 24 2018	&	58201.84	&	23.45	&	...	&	14.29(0.04)	&	13.99(0.02)	&	13.70(0.04)	&	13.47(0.04)	&	...	&	...	&	...	&	KOT	\\
Mar. 25 2018	&	58202.47	&	24.08	&	...	&	14.40(0.04)	&	13.89(0.07)	&	13.68(0.09)	&	13.49(0.07)	&	...	&	...	&	...	&	TNT	\\
Mar. 25 2018	&	58202.67	&	24.28	&	13.75(0.03)	&	14.36(0.01)	&	13.93(0.01)	&	...	&	...	&	14.11(0.01)	&	13.85(0.01)	&	13.87(0.01)	&	LJT	\\
Mar. 26 2018	&	58203.51	&	25.12	&	...	&	...	&	14.01(0.05)	&	13.76(0.09)	&	13.54(0.05)	&	...	&	...	&	...	&	TNT	\\
Mar. 26 2018	&	58203.67	&	25.28	&	13.81(0.03)	&	14.38(0.02)	&	13.94(0.01)	&	...	&	...	&	14.14(0.02)	&	13.86(0.02)	&	13.86(0.02)	&	LJT	\\
Mar. 29 2018	&	58206.67	&	28.28	&	13.95(0.02)	&	14.49(0.01)	&	14.00(0.01)	&	...	&	...	&	14.21(0.01)	&	13.91(0.01)	&	13.89(0.01)	&	LJT	\\
Apr. 01 2018	&	58209.65	&	31.26	&	14.08(0.02)	&	14.55(0.02)	&	14.04(0.02)	&	...	&	...	&	14.27(0.02)	&	13.91(0.02)	&	13.87(0.02)	&	LJT	\\
Apr. 02 2018	&	58210.67	&	32.28	&	14.14(0.03)	&	14.60(0.02)	&	14.05(0.02)	&	...	&	...	&	14.27(0.02)	&	13.92(0.02)	&	13.91(0.01)	&	LJT	\\
Apr. 02 2018	&	58210.79	&	32.40	&	...	&	14.55(0.06)	&	14.11(0.04)	&	13.75(0.08)	&	13.48(0.07)	&	...	&	...	&	...	&	KOT	\\
Apr. 02 2018	&	58210.82	&	32.43	&	...	&	14.61(0.02)	&	14.16(0.01)	&	13.80(0.02)	&	13.49(0.03)	&	...	&	...	&	...	&	KOT	\\
Apr. 03 2018	&	58211.78	&	33.39	&	...	&	14.56(0.05)	&	14.13(0.03)	&	13.77(0.06)	&	13.48(0.06)	&	...	&	...	&	...	&	KOT	\\
Apr. 08 2018	&	58216.48	&	38.09	&	...	&	14.80(0.05)	&	14.12(0.06)	&	13.81(0.04)	&	13.53(0.08)	&	...	&	...	&	...	&	TNT	\\
Apr. 08 2018	&	58216.61	&	38.22	&	14.55(0.03)	&	14.78(0.02)	&	14.18(0.01)	&	...	&	...	&	14.45(0.02)	&	14.01(0.02)	&	13.89(0.03)	&	LJT	\\
Apr. 08 2018	&	58216.80	&	38.41	&	...	&	14.77(0.03)	&	14.22(0.02)	&	13.84(0.05)	&	13.51(0.05)	&	...	&	...	&	...	&	KOT	\\
Apr. 10 2018	&	58218.49	&	40.10	&	...	&	14.88(0.04)	&	14.15(0.04)	&	13.79(0.04)	&	13.54(0.10)	&	...	&	...	&	...	&	TNT	\\
Apr. 11 2018	&	58219.49	&	41.10	&	...	&	14.91(0.06)	&	14.17(0.05)	&	13.83(0.08)	&	13.56(0.08)	&	...	&	...	&	...	&	TNT	\\
Apr. 12 2018	&	58220.67	&	42.28	&	14.70(0.02)	&	14.88(0.02)	&	14.23(0.01)	&	...	&	...	&	14.53(0.01)	&	14.05(0.02)	&	13.99(0.01)	&	LJT	\\
\hline
\hline
\end{tabular}

$^a${Uncertainties (in parentheses) are $1\sigma$.}

$^b${The epoch is relative to the explosion date, MJD = 58178.39.}
\label{Tab:Pho_Ground}
\end{table*}

\begin{table*}
\scriptsize
\caption{Ground-based photometry of SN 2018zd (continued)$^a$}.
\begin{tabular}{lccccccccccc}
\hline\hline
Date (UT)& MJD & Epoch (d)$^b$ & $U$ (mag) & $B$ (mag) & $V$ (mag) & $R$ (mag)& $I$ (mag)& $g$ (mag) & $r$ (mag)& $i$ (mag) & Facility\\
\hline
Apr. 13 2018	&	58221.83	&	43.44	&	...	&	14.97(0.02)	&	14.32(0.02)	&	13.91(0.02)	&	13.58(0.01)	&	...	&	...	&	...	&	KOT	\\
Apr. 14 2018	&	58222.96	&	44.57	&	...	&	14.91(0.04)	&	14.21(0.04)	&	...	&	...	&	...	&	14.08(0.01)	&	13.96(0.04)	&	WT	\\
Apr. 15 2018	&	58223.49	&	45.10	&	...	&	15.04(0.05)	&	14.24(0.04)	&	13.87(0.08)	&	13.58(0.08)	&	...	&	...	&	...	&	TNT	\\
Apr. 16 2018	&	58224.51	&	46.12	&	...	&	15.09(0.04)	&	14.25(0.03)	&	13.89(0.08)	&	13.62(0.09)	&	...	&	...	&	...	&	TNT	\\
Apr. 17 2018	&	58225.56	&	47.17	&	14.98(0.02)	&	15.05(0.03)	&	14.27(0.02)	&	...	&	...	&	14.65(0.03)	&	14.11(0.03)	&	13.96(0.03)	&	LJT	\\
Apr. 18 2018	&	58226.49	&	48.10	&	...	&	...	&	14.27(0.05)	&	13.89(0.09)	&	13.58(0.07)	&	...	&	...	&	...	&	TNT	\\
Apr. 12 2018	&	58226.81	&	48.42	&	...	&	15.13(0.03)	&	14.38(0.02)	&	13.95(0.05)	&	13.60(0.05)	&	...	&	...	&	...	&	KOT	\\
Apr. 18 2018	&	58226.91	&	48.52	&	...	&	15.02(0.04)	&	14.27(0.04)	&	...	&	...	&	...	&	14.13(0.02)	&	14.01(0.04)	&	WT	\\
Apr. 20 2018	&	58228.51	&	50.12	&	...	&	...	&	14.32(0.05)	&	13.91(0.06)	&	13.61(0.10)	&	...	&	...	&	...	&	TNT	\\
Apr. 21 2018	&	58229.70	&	51.31	&	15.22(0.03)	&	15.16(0.03)	&	14.36(0.01)	&	...	&	...	&	14.71(0.01)	&	14.12(0.01)	&	14.05(0.02)	&	LJT	\\
Apr. 22 2018	&	58230.82	&	52.43	&	...	&	15.25(0.02)	&	14.43(0.02)	&	13.99(0.03)	&	13.64(0.03)	&	...	&	...	&	...	&	KOT	\\
Apr. 25 2018	&	58233.53	&	55.14	&	15.44(0.02)	&	15.28(0.02)	&	14.43(0.01)	&	...	&	...	&	14.82(0.02)	&	14.20(0.01)	&	14.08(0.01)	&	LJT	\\
Apr. 27 2018	&	58235.56	&	57.17	&	15.48(0.04)	&	15.31(0.03)	&	14.41(0.02)	&	...	&	...	&	14.82(0.01)	&	14.19(0.01)	&	14.07(0.02)	&	LJT	\\
Apr. 27 2018	&	58235.80	&	57.41	&	...	&	15.38(0.01)	&	14.53(0.03)	&	14.04(0.02)	&	13.69(0.01)	&	...	&	...	&	...	&	KOT	\\
Apr. 28 2018	&	58236.81	&	58.42	&	...	&	15.42(0.07)	&	14.51(0.03)	&	14.04(0.06)	&	13.70(0.03)	&	...	&	...	&	...	&	KOT	\\
Apr. 30 2018	&	58238.82	&	60.43	&	...	&	15.43(0.05)	&	14.54(0.02)	&	14.07(0.05)	&	13.70(0.06)	&	...	&	...	&	...	&	KOT	\\
Mar. 02 2018	&	58240.80	&	62.41	&	...	&	15.53(0.06)	&	14.55(0.04)	&	14.08(0.06)	&	13.72(0.05)	&	...	&	...	&	...	&	KOT	\\
May 03 2018	&	58241.51	&	63.12	&	...	&	15.53(0.05)	&	14.48(0.04)	&	14.05(0.08)	&	13.71(0.07)	&	...	&	...	&	...	&	TNT	\\
May 04 2018	&	58242.52	&	64.13	&	...	&	15.57(0.06)	&	14.52(0.05)	&	14.06(0.04)	&	13.70(0.09)	&	...	&	...	&	...	&	TNT	\\
May 07 2018	&	58245.53	&	67.14	&	...	&	15.60(0.05)	&	14.54(0.04)	&	14.04(0.09)	&	13.65(0.10)	&	...	&	...	&	...	&	TNT	\\
May 08 2018	&	58246.56	&	68.17	&	15.99(0.04)	&	15.58(0.01)	&	14.60(0.01)	&	...	&	...	&	15.04(0.01)	&	14.32(0.01)	&	14.20(0.01)	&	LJT	\\
May 13 2018	&	58250.87	&	72.48	&	...	&	...	&	14.61(0.06)	&	...	&	...	&	...	&	14.36(0.02)	&	14.19(0.05)	&	WT	\\
May 14 2018	&	58251.86	&	73.47	&	...	&	...	&	14.63(0.05)	&	...	&	...	&	...	&	14.38(0.02)	&	14.21(0.03)	&	WT	\\
May 15 2018	&	58253.54	&	75.15	&	16.23(0.05)	&	15.75(0.02)	&	14.69(0.01)	&	...	&	...	&	15.18(0.01)	&	14.37(0.02)	&	14.29(0.02)	&	LJT	\\
May 20 2018	&	58257.86	&	79.47	&	...	&	15.72(0.06)	&	14.70(0.06)	&	...	&	...	&	...	&	14.43(0.02)	&	14.25(0.04)	&	WT	\\
May 22 2018	&	58259.87	&	81.48	&	...	&	15.76(0.06)	&	14.74(0.05)	&	...	&	...	&	...	&	14.46(0.02)	&	14.29(0.04)	&	WT	\\
Mar. 23 2018	&	58261.98	&	83.59	&	...	&	16.00(0.02)	&	14.87(0.02)	&	14.33(0.02)	&	13.92(0.02)	&	...	&	...	&	...	&	KOT	\\
May 27 2018	&	58264.89	&	86.50	&	...	&	15.87(0.07)	&	14.82(0.05)	&	...	&	...	&	...	&	14.52(0.02)	&	14.34(0.04)	&	WT	\\
Jun. 06 2018	&	58275.88	&	97.49	&	...	&	16.18(0.08)	&	15.03(0.06)	&	...	&	...	&	...	&	14.69(0.03)	&	14.49(0.04)	&	WT	\\
Jun. 07 2018	&	58276.88	&	98.49	&	...	&	16.18(0.07)	&	15.04(0.07)	&	...	&	...	&	...	&	14.70(0.04)	&	14.49(0.04)	&	WT	\\
Jun. 07 2018	&	58277.00	&	98.61	&	...	&	16.37(0.03)	&	15.12(0.03)	&	14.54(0.02)	&	14.12(0.02)	&	...	&	...	&	...	&	KOT	\\
Jun. 09 2018	&	58278.96	&	100.57	&	...	&	16.42(0.05)	&	15.15(0.03)	&	14.61(0.03)	&	14.17(0.07)	&	...	&	...	&	...	&	KOT	\\
Jun. 11 2018	&	58281.01	&	102.62	&	...	&	16.53(0.03)	&	15.22(0.02)	&	14.62(0.03)	&	14.19(0.07)	&	...	&	...	&	...	&	KOT	\\
Jun. 14 2018	&	58283.89	&	105.50	&	...	&	16.37(0.09)	&	15.21(0.06)	&	...	&	...	&	...	&	14.82(0.02)	&	14.62(0.04)	&	WT	\\
Jun. 16 2018	&	58285.88	&	107.49	&	...	&	16.46(0.07)	&	15.27(0.07)	&	...	&	...	&	...	&	14.89(0.02)	&	...	&	WT	\\
Jun. 16 2018	&	58286.00	&	107.61	&	...	&	16.71(0.03)	&	15.35(0.01)	&	14.73(0.03)	&	14.31(0.04)	&	...	&	...	&	...	&	KOT	\\
Jul. 13 2018	&	58312.97	&	134.58	&	...	&	19.41(0.05)	&	18.14(0.03)	&	17.13(0.02)	&	16.70(0.02)	&	...	&	...	&	...	&	KOT	\\
Jul. 15 2018	&	58314.89	&	136.50	&	...	&	19.18(0.18)	&	...	&	...	&	...	&	...	&	17.35(0.04)	&	17.02(0.06)	&	WT	\\
Jul. 16 2018	&	58315.98	&	137.59	&	...	&	19.47(0.08)	&	18.17(0.03)	&	17.24(0.02)	&	16.75(0.02)	&	...	&	...	&	...	&	KOT	\\
Jul. 26 2018	&	58326.00	&	147.61	&	...	&	19.51(0.27)	&	18.23(0.07)	&	17.18(0.04)	&	16.78(0.03)	&	...	&	...	&	...	&	KOT	\\
Aug. 03 2018	&	58333.01	&	154.62	&	...	&	19.50(0.10)	&	18.37(0.04)	&	17.36(0.02)	&	16.95(0.04)	&	...	&	...	&	...	&	KOT	\\
Aug. 08 2018	&	58338.99	&	160.60	&	...	&	19.81(0.07)	&	18.39(0.03)	&	17.37(0.02)	&	17.03(0.03)	&	...	&	...	&	...	&	KOT	\\
Aug. 10 2018	&	58340.85	&	162.46	&	...	&	19.61(0.08)	&	18.47(0.03)	&	17.44(0.03)	&	17.06(0.04)	&	...	&	...	&	...	&	KOT	\\
Aug. 17 2018	&	58347.85	&	169.46	&	...	&	20.22(0.51)	&	19.06(0.17)	&	17.63(0.06)	&	17.08(0.10)	&	...	&	...	&	...	&	KOT	\\
Aug. 18 2018	&	58348.83	&	170.44	&	...	&	20.06(0.32)	&	18.83(0.10)	&	17.43(0.05)	&	17.09(0.06)	&	...	&	...	&	...	&	KOT	\\
Aug. 08 2018	&	58348.86	&	170.47	&	...	&	19.68(0.05)	&	18.49(0.02)	&	...	&	...	&	18.98(0.02)	&	17.50(0.02)	&	17.71(0.01)	&	LJT	\\
Aug. 21 2018	&	58351.86	&	173.47	&	...	&	19.25(0.16)	&	18.38(0.08)	&	17.46(0.05)	&	17.03(0.04)	&	...	&	...	&	...	&	KOT	\\
Aug. 30 2018	&	58360.89	&	182.50	&	...	&	19.67(0.03)	&	18.63(0.02)	&	...	&	...	&	19.10(0.02)	&	17.64(0.02)	&	17.83(0.02)	&	LJT	\\
Aug. 30 2018	&	58360.90	&	182.51	&	...	&	19.55(0.37)	&	18.81(0.11)	&	17.58(0.05)	&	17.23(0.04)	&	...	&	...	&	...	&	KOT	\\
Sep. 06 2018	&	58367.76	&	189.37	&	...	&	19.93(0.11)	&	18.68(0.05)	&	17.63(0.03)	&	17.25(0.04)	&	...	&	...	&	...	&	KOT	\\
Sep. 11 2018	&	58372.78	&	194.39	&	...	&	20.07(0.09)	&	18.68(0.07)	&	17.76(0.03)	&	17.13(0.04)	&	...	&	...	&	...	&	KOT	\\
Sep. 15 2018	&	58377.01	&	198.62	&	...	&	20.18(0.10)	&	18.78(0.03)	&	17.71(0.01)	&	17.33(0.02)	&	...	&	...	&	...	&	KOT	\\
Sep. 20 2018	&	58382.12	&	203.73	&	...	&	19.81(0.37)	&	18.90(0.10)	&	17.67(0.04)	&	17.42(0.03)	&	...	&	...	&	...	&	KOT	\\
Sep. 24 2018	&	58386.00	&	207.61	&	...	&	19.94(0.42)	&	18.86(0.14)	&	17.71(0.04)	&	17.38(0.03)	&	...	&	...	&	...	&	KOT	\\
Oct. 11 2018	&	58402.71	&	224.32	&	...	&	19.81(0.11)	&	19.04(0.06)	&	17.87(0.03)	&	17.65(0.03)	&	...	&	...	&	...	&	KOT	\\
Oct. 14 2018	&	58405.07	&	226.68	&	...	&	20.01(0.06)	&	18.91(0.04)	&	17.94(0.02)	&	17.66(0.03)	&	...	&	...	&	...	&	KOT	\\
Oct. 15 2018	&	58406.76	&	228.37	&	...	&	20.02(0.05)	&	18.96(0.02)	&	...	&	...	&	19.41(0.03)	&	18.03(0.02)	&	18.28(0.02)	&	LJT	\\
Oct. 18 2018	&	58409.76	&	231.37	&	...	&	20.12(0.05)	&	19.08(0.01)	&	...	&	...	&	19.45(0.01)	&	18.05(0.03)	&	18.31(0.03)	&	LJT	\\
Nov. 03 2018	&	58425.76	&	247.37	&	...	&	20.18(0.05)	&	19.26(0.03)	&	...	&	...	&	19.57(0.02)	&	18.23(0.02)	&	18.50(0.03)	&	LJT	\\
Nov. 04 2018	&	58426.79	&	248.40	&	...	&	20.10(0.05)	&	19.13(0.03)	&	...	&	...	&	19.52(0.02)	&	18.25(0.02)	&	18.49(0.03)	&	LJT	\\
Nov. 05  2018	&	58427.04	&	248.65	&	...	&	20.25(0.16)	&	19.23(0.10)	&	18.17(0.06)	&	18.18(0.06)	&	...	&	...	&	...	&	KOT	\\
Nov. 12 2018	&	58434.68	&	256.29	&	...	&	20.03(0.19)	&	19.35(0.09)	&	18.19(0.04)	&	17.99(0.05)	&	...	&	...	&	...	&	KOT	\\
Nov. 14 2018	&	58436.76	&	258.37	&	...	&	20.30(0.05)	&	19.34(0.02)	&	...	&	...	&	19.65(0.04)	&	18.33(0.04)	&	18.45(0.05)	&	LJT	\\
Nov. 18 2018	&	58440.92	&	262.53	&	...	&	20.33(0.06)	&	19.22(0.04)	&	...	&	...	&	19.65(0.03)	&	18.35(0.03)	&	18.54(0.04)	&	LJT	\\
Nov. 20 2018	&	58442.68	&	264.29	&	...	&	20.83(0.21)	&	19.70(0.11)	&	...	&	...	&	19.97(0.11)	&	18.44(0.04)	&	18.65(0.05)	&	LJT	\\
Nov. 27 2018	&	58449.76	&	271.37	&	...	&	20.33(0.06)	&	19.44(0.03)	&	...	&	...	&	19.83(0.04)	&	18.50(0.03)	&	18.70(0.05)	&	LJT	\\
Dec. 04 2018	&	58456.75	&	278.36	&	...	&	20.31(0.06)	&	19.35(0.03)	&	...	&	...	&	19.78(0.02)	&	18.55(0.02)	&	18.74(0.03)	&	LJT	\\
Dec. 08 2018	&	58460.74	&	282.35	&	...	&	20.47(0.04)	&	19.57(0.02)	&	...	&	...	&	19.85(0.02)	&	18.59(0.03)	&	18.71(0.04)	&	LJT	\\
Dec. 31 2018	&	58483.70	&	305.31	&	...	&	...	&	...	&	...	&	...	&	20.04(0.02)	&	18.89(0.03)	&	19.02(0.04)	&	LJT	\\
Jan. 09 2019	&	58492.68	&	314.29	&	...	&	20.67(0.09)	&	19.91(0.07)	&	...	&	...	&	20.08(0.05)	&	19.00(0.04)	&	19.02(0.04)	&	LJT	\\
Jan. 27 2019	&	58510.59	&	332.20	&	...	&	20.71(0.05)	&	20.01(0.04)	&	...	&	...	&	20.18(0.04)	&	19.12(0.03)	&	19.26(0.04)	&	LJT	\\
Feb. 02 2019	&	58516.59	&	338.20	&	...	&	20.93(0.06)	&	20.07(0.03)	&	...	&	...	&	20.19(0.03)	&	19.15(0.04)	&	19.19(0.04)	&	LJT	\\
Mar. 02 2019	&	58544.53	&	366.14	&	...	&	21.18(0.08)	&	20.19(0.05)	&	...	&	...	&	20.56(0.05)	&	19.72(0.05)	&	19.75(0.05)	&	LJT	\\
Mar. 29 2019	&	58571.53	&	393.14	&	...	&	...	&	...	&	...	&	...	&	20.64(0.07)	&	19.83(0.06)	&	19.84(0.07)	&	LJT	\\
Apr. 07 2019	&	58580.52	&	402.13	&	...	&	...	&	...	&	...	&	...	&	...	&	19.94(0.08)	&	19.89(0.07)	&	LJT	\\
Apr. 23 2019	&	58596.53	&	418.14	&	...	&	...	&	...	&	...	&	...	&	20.99(0.12)	&	20.14(0.06)	&	20.11(0.07)	&	LJT	\\
May 31 2019	&	58634.54	&	456.15	&	...	&	...	&	...	&	...	&	...	&	...	&	20.51(0.18)	&	20.45(0.13)	&	LJT	\\
\hline
\hline
\end{tabular}
$^a${Uncertainties (in parentheses) are $1\sigma$.}

$^b${The epoch is relative to the explosion date, MJD = 58178.39.}
\label{Tab:Pho_Ground2}
\end{table*}

\begin{table*}
\caption{{\it Swift} UVOT photometry of SN 2018zd (Vega magnitude system)$^a$}
\scriptsize
\begin{tabular}{lcccccccc}
\hline\hline
Date (UT)& MJD& Epoch (d)$^b$ & $uvw2$ (mag) & $uvm2$ (mag) & $uvw1$ (mag) & $u$ (mag) & $b$ (mag) & $v$ (mag)\\
\hline
Mar. 04 2018	&	58181.29	&	2.90	&	16.00(0.04)	&	15.86(0.03)	&	15.53(0.04)	&	15.17(0.04)	&	16.17(0.04)	&	15.99(0.06)	\\
Mar. 04 2018	&	58181.35	&	2.96	&	15.71(0.04)	&	15.68(0.06)	&	15.34(0.04)	&	15.01(0.04)	&	16.06(0.04)	&	15.82(0.06)	\\
Mar. 05 2018	&	58182.35	&	3.96	&	14.10(0.02)	&	14.19(0.02)	&	14.03(0.02)	&	13.92(0.03)	&	15.04(0.03)	&	15.01(0.04)	\\
Mar. 06 2018	&	58183.21	&	4.82	&	13.48(0.02)	&	13.49(0.06)	&	13.47(0.03)	&	13.43(0.03)	&	14.61(0.03)	&	14.56(0.04)	\\
Mar. 07 2018	&	58184.28	&	5.89	&	13.07(0.11)	&	13.18(0.03)	&	13.25(0.11)	&	12.96(0.04)	&	14.21(0.04)	&	14.24(0.07)	\\
Mar. 08 2018	&	58185.14	&	6.75	&	12.73(0.02)	&	12.88(0.03)	&	12.79(0.07)	&	12.69(0.03)	&	13.88(0.03)	&	13.84(0.04)	\\
Mar. 10 2018	&	58187.23	&	8.84	&	12.86(0.02)	&	12.89(0.02)	&	12.77(0.07)	&	12.55(0.03)	&	13.71(0.02)	&	13.61(0.03)	\\
Mar. 12 2018	&	58189.42	&	11.03	&	13.47(0.02)	&	13.28(0.02)	&	13.13(0.07)	&	12.78(0.03)	&	13.85(0.02)	&	13.68(0.03)	\\
Mar. 14 2018	&	58191.74	&	13.35	&	13.85(0.02)	&	13.58(0.02)	&	13.28(0.07)	&	12.89(0.03)	&	13.92(0.03)	&	13.66(0.03)	\\
Mar. 16 2018	&	58193.53	&	15.14	&	14.33(0.03)	&	13.99(0.03)	&	13.53(0.07)	&	13.02(0.03)	&	14.00(0.03)	&	13.75(0.03)	\\
Mar. 18 2018	&	58195.85	&	17.46	&	14.56(0.02)	&	14.20(0.03)	&	13.76(0.07)	&	13.11(0.03)	&	14.05(0.02)	&	13.80(0.03)	\\
Mar. 20 2018	&	58197.72	&	19.33	&	14.83(0.03)	&	14.49(0.03)	&	13.92(0.07)	&	13.22(0.03)	&	14.14(0.03)	&	13.86(0.03)	\\
Mar. 30 2018	&	58207.34	&	28.95	&	16.01(0.04)	&	16.07(0.04)	&	15.03(0.07)	&	13.85(0.03)	&	14.41(0.03)	&	14.04(0.04)	\\
Apr. 01 2018	&	58209.83	&	31.44	&	16.42(0.05)	&	16.55(0.05)	&	15.35(0.08)	&	13.98(0.03)	&	14.50(0.03)	&	14.08(0.04)	\\
Apr. 05 2018	&	58213.41	&	35.02	&	17.01(0.05)	&	17.23(0.05)	&	15.77(0.08)	&	14.25(0.03)	&	14.63(0.03)	&	14.12(0.03)	\\
Apr. 07 2018	&	58215.18	&	36.79	&	17.21(0.06)	&	17.71(0.07)	&	16.01(0.08)	&	14.38(0.03)	&	14.72(0.03)	&	14.21(0.04)	\\
Apr. 19 2018	&	58227.13	&	48.74	&	18.16(0.15)	&	19.20(0.26)	&	17.16(0.14)	&	15.44(0.06)	&	14.96(0.04)	&	14.29(0.05)	\\
Apr. 24 2018	&	58232.54	&	54.15	&	18.46(0.10)	&	19.18(0.14)	&	17.64(0.11)	&	15.64(0.04)	&	15.15(0.03)	&	14.38(0.03)	\\
Apr. 30 2018	&	58238.74	&	60.35	&	18.80(0.11)	&	19.38(0.14)	&	17.82(0.11)	&	16.00(0.05)	&	15.34(0.03)	&	14.44(0.03)	\\
May 06 2018	&	58244.72	&	66.33	&	18.87(0.12)	&	19.37(0.15)	&	18.14(0.12)	&	16.42(0.06)	&	15.50(0.03)	&	14.52(0.03)	\\
May 12 2018	&	58250.25	&	71.86	&	18.79(0.11)	&	19.67(0.17)	&	18.31(0.13)	&	16.65(0.07)	&	15.61(0.03)	&	14.62(0.04)	\\
May 18 2018	&	58256.55	&	78.16	&	19.29(0.17)	&	19.81(0.22)	&	18.64(0.16)	&	16.81(0.08)	&	15.73(0.03)	&	14.72(0.04)	\\
May 24 2018	&	58262.92	&	84.53	&	19.24(0.19)	&	20.02(0.26)	&	18.64(0.17)	&	17.20(0.11)	&	15.93(0.04)	&	14.82(0.04)	\\
Jun. 01 2018	&	58270.16	&	91.77	&	19.41(0.18)	&	...	&	18.23(0.18)	&	17.28(0.10)	&	16.16(0.03)	&	14.97(0.04)	\\
Jun. 09 2018	&	58278.51	&	100.12	&	19.75(0.24)	&	...	&	18.92(0.21)	&	17.60(0.12)	&	16.23(0.04)	&	15.08(0.04)	\\
Jun. 13 2018	&	58282.26	&	103.87	&	19.36(0.18)	&	...	&	19.36(0.18)	&	17.84(0.11)	&	16.44(0.04)	&	15.21(0.04)	\\
Jun. 19 2018	&	58288.23	&	109.84	&	19.84(0.28)	&	...	&	18.92(0.35)	&	18.31(0.14)	&	16.74(0.04)	&	15.35(0.04)	\\
Jun. 25 2018	&	58294.34	&	115.95	&	19.89(0.29)	&	...	&	19.24(0.30)	&	18.26(0.18)	&	16.78(0.12)	&	15.61(0.05)	\\	
\hline
\hline
\end{tabular}

$^a${Uncertainties (in parentheses) are $1\sigma$.}

$^b${The epoch is relative to the explosion date, MJD = 58178.39.}
\label{Tab:Swiftpho}
\end{table*}



\begin{table*}
\caption{Journal of spectroscopic observations of SN 2018zd}
\scriptsize
\begin{tabular}{lccccccc}
\hline\hline
Date (UT) & MJD & Epoch (d)$^a$ & Range (\AA) & Disp. (\AA\ pix$^{-1}$) & Exp (s) &airmass & Telescope+Inst.\\
\hline
Mar. 02, 2018	&	58179.61	&	1.22	&	3500--8900	&	2.85	&	2100	&	1.64	&	LJT+YFOSC	\\
Mar. 04, 2018	&	58181.74	&	3.35	&	3500--8900	&	2.85	&	750	&	2.03	&	LJT+YFOSC	\\
Mar. 05, 2018	&	58182.74	&	4.35	&	3500--8900	&	2.85	&	2100	&	2.00	&	LJT+YFOSC	\\
Mar. 07, 2018	&	58184.69	&	6.30	&	3500--8900	&	2.85	&	1500	&	1.83	&	LJT+YFOSC	\\
Mar. 09, 2018	&	58186.52	&	8.13	&	3960--8820	&	2.78	&	2400	&	1.30	&	XLT+BFOSC	\\
Mar. 09, 2018	&	58186.71	&	8.32	&	3500--8900	&	2.85	&	1200	&	1.93	&	LJT+YFOSC	\\
Mar. 10, 2018	&	58187.57	&	9.18	&	3500--8900	&	2.85	&	1500	&	1.62	&	LJT+YFOSC	\\
Mar. 11, 2018	&	58188.68	&	10.29	&	3500--8900	&	2.85	&	1500	&	1.84	&	LJT+YFOSC	\\
Mar. 13, 2018	&	58190.47	&	12.08	&	3980--8830	&	2.78	&	2100	&	1.27	&	XLT+BFOSC	\\
Mar. 13, 2018	&	58190.62	&	12.23	&	3500--8900	&	2.85	&	1800	&	1.71	&	LJT+YFOSC	\\
Mar. 15, 2018	&	58192.68	&	14.29	&	3500--8900	&	2.85	&	1500	&	1.87	&	LJT+YFOSC	\\
Mar. 18, 2018	&	58195.74	&	17.35	&	3500--8900	&	2.85	&	1500	&	2.18	&	LJT+YFOSC	\\
Mar. 20, 2018	&	58197.68	&	19.29	&	3500--8900	&	2.85	&	1500	&	1.94	&	LJT+YFOSC	\\
Mar. 22, 2018	&	58199.53	&	21.14	&	3500--8900	&	2.85	&	1500	&	1.62	&	LJT+YFOSC	\\
Mar. 24, 2018	&	58201.47	&	23.08	&	3960--8820	&	2.78	&	2400	&	1.29	&	XLT+BFOSC	\\
Mar. 24, 2018	&	58201.71	&	23.32	&	3500--8900	&	2.85	&	1500	&	2.12	&	LJT+YFOSC	\\
Mar. 29, 2018	&	58206.67&	28.28	&	3500-8900	&	2.85	&	1800	&	2.04	&	LJT+YFOSC	\\
Apr. 02, 2018	&	58210.69	&	32.30	&	3500--8900	&	2.85	&	1800	&	2.11	&	LJT+YFOSC	\\
Apr. 13, 2018	&	58221.63	&	43.24	&	3500--8900	&	2.85	&	1300	&	2.01	&	LJT+YFOSC	\\
Apr. 15, 2018	&	58223.53	&	45.14	&	3960--8820	&	2.78	&	2400	&	1.44	&	XLT+BFOSC	\\
Apr. 23, 2018	&	58231.50	&	53.11	&	3450--8780	&	2.78	&	2700	&	1.42	&	XLT+BFOSC	\\
Apr. 23, 2018	&	58231.68	&	53.29	&	3500--8900	&	2.85	&	1350	&	2.39	&	LJT+YFOSC	\\
Apr. 27, 2018	&	58235.54	&	57.15	&	3960--8820	&	2.78	&	2400	&	1.53	&	XLT+BFOSC	\\
Apr. 28, 2018	&	58236.51	&	58.12	&	3960--8820	&	2.78	&	2400	&	1.45	&	XLT+BFOSC	\\
May 08, 2018	&	58246.50	&	68.11	&	4090--8800	&	2.78	&	2700	&	1.50	&	XLT+BFOSC	\\
Aug. 30, 2018	&	58360.87	&	182.48	&	3500--8900	&	2.85	&	2100	&	1.96	&	LJT+YFOSC	\\
Oct. 15, 2018	&	58406.73	&	228.34	&	3500--8900	&	2.85	&	2100	&	2.00	&	LJT+YFOSC	\\
Oct. 30, 2018	&	58421.80	&	243.41	&	3980--8830	&	2.78	&	3000	&	1.27	&	XLT+BFOSC	\\
Nov. 04, 2018	&	58426.18	&	247.79	&	5280--9300	&	2.29	&	1800	&	1.91	&	APO+DIS	\\
Nov. 06, 2018	&	58428.71	&	250.32	&	3500--8900	&	2.85	&	2700	&	1.87	&	LJT+YFOSC	\\
Nov. 18, 2018	&	58440.86	&	262.47	&	3860--8680	&	2.78	&	3600	&	1.33	&	XLT+BFOSC	\\
Dec. 10, 2018	&	58462.40	&	284.01	&	5260--9400	&	2.30	&	1800	&	1.45	&	APO+DIS	\\
Dec. 17, 2018	&	58469.79	&	291.40	&	3970--8830	&	2.78	&	3600	&	1.34	&	XLT+BFOSC	\\
Nov. 01, 2019	&	58484.59	&	306.20	&	3850--8680	&	2.78	&	3600	&	1.28	&	XLT+BFOSC	\\
Jan. 10, 2019	&	58493.61	&	315.22	&	3500--8900	&	2.85	&	3000	&	1.68	&	LJT+YFOSC	\\
Feb. 02, 2019	&	58516.62	&	338.23	&	3500--8900	&	2.85	&	3600	&	1.68	&	LJT+YFOSC	\\
Apr. 04, 2019	&	58577.27	&	398.88	&	3110--10300	&	0.91	&	940	&	2.06	&	Keck I+LRIS$^b$	\\
\hline
\hline
\end{tabular}

$^a${The epoch is relative to the explosion date, MJD = 58178.39.}

$^b${Blue and red sides combined; parameters listed are the average values.}
\label{Tab:Spec_log}
\end{table*}

\begin{table}
\caption{Clear-band photometry of SN\,2018zd (``Bright Supernovae'')$^a$}
\scriptsize
\centering
\begin{tabular}{lcccc}
\hline\hline
Date (UT) & MJD & Epoch (d)$^b$ & Mag & Observer\\
\hline
Mar. 01.54, 2018	&	58178.54	&	0.15	&	18.0	&	Koichi Itagaki 	\\
Mar.  02.49, 2018	&	58179.49	&	1.10	&	17.8	&	Koichi Itagaki	\\
Mar. 03.49, 2018	&	58180.64	&	2.25	&	16.8	&	Koichi Itagaki	\\
Mar. 03.64  2018	&	58180.79	&	2.40	&	16.5	&	Koichi Itagaki	\\
Mar. 03.79, 2018	&	58180.94	&	2.55	&	16.6	&	Koichi Itagaki	\\
Mar. 04.53, 2018	&	58181.68	&	3.29	&	15.5	&	Yuji Tanaka	\\
Mar. 06.54, 2018	&	58183.69	&	5.30	&	14.2	&	Yuji Tanaka	\\
Mar.  06.71,  2018	&	58183.86	&	5.47	&	14.4	&	Koichi Itagaki	\\
Mar. 14.45, 2018	&	58191.60	&	13.21	&	13.6	&	Yuji Tanaka	\\
\hline
\hline
\end{tabular}

$^a${www.rochesterastronomy.org/sn2018/sn2018zd.html; uncertainties are $\sim 0.1$\,mag, considering the translation between the clear band and the $r$ band.}

$^b${The epoch is relative to the explosion date, MJD = 58178.39.}
\label{Tab:LC_ama}
\end{table}



\bsp	
\label{lastpage}
\end{document}